\documentclass[apj]{emulateapj}
\usepackage{apjfonts,ifthen,graphicx,times,lscape}

\newcommand{\etal}{et al.\ }
\newcommand{\swift}{\textit{Swift}}
\newcommand{\swiftbat}{\textit{Swift}-BAT}
\newcommand{\asca}{ASCA}
\newcommand{\rosat}{ROSAT}
\newcommand{\xmm}{\textit{XMM-Newton}}
\newcommand{\chandra}{\textit{Chandra}}
\newcommand{\integral}{INTEGRAL}

\newcommand{\nh}{$n_{\rm H}$}

\newcommand{\batcelldetect}{\texttt{batcelldetect}}
\newcommand{\degree}{\ensuremath{^{\circ}}}

\begin{document}

\title{The 22-Month \swiftbat\ All-Sky Hard X-Ray Survey}

\shorttitle{SWIFT-BAT 22 MONTH HARD X-RAY SURVEY}
\shortauthors{TUELLER ET AL.}
%% You can insert a short comment on the title page using the command below.
%\slugcomment{(accepted by the Astrophysical Journal Supplement)}
% The emulateapj command is:
%\submitted{Draft of 13 March 2009 (to be submitted to the ApJ Supplement)}
\journalinfo{Accepted by the Astrophysical Journal Supplement}

\author{
J.~Tueller\altaffilmark{1},
W.~H.~Baumgartner\altaffilmark{1,2,12,18},
C.~B.~Markwardt\altaffilmark{1,3,12},
G.~K.~Skinner\altaffilmark{1,3,12},
R.~F.~Mushotzky\altaffilmark{1},
M.~Ajello\altaffilmark{4},
S.~Barthelmy\altaffilmark{1},
A.~Beardmore\altaffilmark{5},
W.~N.~Brandt\altaffilmark{7},
D.~Burrows\altaffilmark{7},
G.~Chincarini\altaffilmark{8},
S.~Campana\altaffilmark{8},
J.~Cummings\altaffilmark{1},
G.~Cusumano\altaffilmark{10},
P.~Evans\altaffilmark{5},
E.~Fenimore\altaffilmark{11},
N.~Gehrels\altaffilmark{1},
O.~Godet\altaffilmark{5},
D.~Grupe\altaffilmark{7},
S.~Holland\altaffilmark{1,12},
J.~Kennea\altaffilmark{7},
H.~A.~Krimm\altaffilmark{1,12},
M.~Koss\altaffilmark{3,1,12},
A.~Moretti\altaffilmark{8},
K.~Mukai\altaffilmark{1,2,12},
J.~P.~Osborne\altaffilmark{5},
T.~Okajima\altaffilmark{1,13},
C.~Pagani\altaffilmark{7},
K.~Page\altaffilmark{5},
D.~Palmer\altaffilmark{11},
A.~Parsons\altaffilmark{1},
D.~P.~Schneider\altaffilmark{7},
T.~Sakamoto\altaffilmark{1,14},
R.~Sambruna\altaffilmark{1},
G.~Sato\altaffilmark{17},
M.~Stamatikos\altaffilmark{1,14},
M.~Stroh\altaffilmark{7},
T.N.~Ukwatta\altaffilmark{1,15},
L.~Winter\altaffilmark{16}
}

\altaffiltext{1}{NASA/Goddard Space Flight Center, Astrophysics
 Science Division, Greenbelt, MD 20771}
\altaffiltext{2}{Joint Center for Astrophysics, University of
 Maryland Baltimore County, Baltimore, MD 21250}
\altaffiltext{3}{Department of Astronomy, University of
 Maryland College Park, College Park, MD 20742}
\altaffiltext{4}{SLAC National Laboratory and Kavli Institute for
  Particle Astrophysics and Cosmology, 2575 Sand Hill Road, Menlo
  Park, CA 94025}
\altaffiltext{5}{X-Ray and Observational Astronomy Group/ Department
  of Physics and Astronomy, University of Leicester, Leicester, LE1
  7RH, United Kingdom}
%\altaffiltext{6}{Mullard Space Science Laboratory (MSSL)/ Department
%  of Space \& Climate Physics, University College London, Dorking,
%  United Kingdom}
\altaffiltext{7}{Department of Astronomy \& Astrophysics, The
  Pennsylvania State University, 525 Davey Lab, University Park, PA
  16802}
\altaffiltext{8}{Osservatorio Astronomico di Brera(OAB)/ Istituto
  Nazionale di Astrofisica (INAF), 20121 Milano, Italy}
\altaffiltext{9}{ASDC/ ESA of ESRIN, Via Galileo Galilei, 00044
  Frascati (RM), Italy}
\altaffiltext{10}{IASF-Palermo/ Istituto di Astrofisica Spaziale e
  Fisica Cosmica di Palermo/ Istituto Nazionale di Astrofisica (INAF),
  90146 Palermo, Italy} 
\altaffiltext{11}{LANL/Los Alamos National Laboratory, Los Alamos, NM 87545}
\altaffiltext{12}{CRESST/ Center for Research and Exploration in Space
  Science and Technology, 10211 Wincopin Circle, Suite 500, Columbia,
  MD 21044}
\altaffiltext{13}{Department of Physics \& Astronomy, The Johns
  Hopkins University, 3400 North Charles Street Baltimore, Maryland
  21218}
\altaffiltext{14}{Oak Ridge Associated Universities (ORAU), OAB-44,
  P.O. Box 117 Oak Ridge, TN 37831}
\altaffiltext{15}{Department of Physics/ The George Washington
  University (GWU), 2121 I Street, N.W., Washington, DC 20052}
\altaffiltext{16}{Center for Astrophysics and Space Astronomy,
  University of Colorado, 389 UCB, Boulder, CO 80309 }
\altaffiltext{17}{Institute of Space and Astronautical Science, JAXA,
  Kanagawa 229-8510, Japan}
\altaffiltext{18}{Corresponding author: Wayne.Baumgartner@nasa.gov}

\begin{abstract}

We present the catalog of sources detected in the first 22 months of
data from the hard X-ray survey (14--195~keV) conducted with the BAT
coded mask imager on the \swift\ satellite.  The catalog contains 461
sources detected above the $4.8\sigma$ level with BAT.  High angular
resolution X-ray data for every source from \swift\ XRT or archival
data have allowed associations to be made with known counterparts in
other wavelength bands for over 97\% of the detections, including the
discovery of $\sim30$ galaxies previously unknown as AGN and several
new Galactic sources.  A total of 266 of the sources are associated
with Seyfert galaxies (median redshift $z\sim0.03$) or blazars, with
the majority of the remaining sources associated with X-ray binaries
in our Galaxy.  This ongoing survey is the first uniform all sky hard
X-ray survey since HEAO-1 in 1977.

Since the publication of the 9-month BAT survey we have increased the
number of energy channels from 4 to 8 and have substantially increased
the number of sources with accurate average spectra.  The BAT 22-month
catalog is the product of the most sensitive all-sky survey in the
hard X-ray band, with a detection sensitivity (4.8$\sigma$) of
$2.2\times10^{-11}$~erg~cm$^{-2}$~s$^{-1}$ (1~mCrab) over most of the
sky in the 14--195~keV band.

\end{abstract}


%\tableofcontents
%\listoffigures
%\listoftables

%--------------------------------------------------------------------
\section{Introduction}

Surveys of the whole sky which are complete to a well-defined
threshold not only provide a basis for statistical population studies
but are also a vehicle for the discovery of new phenomena.  Compared
with lower X-ray energies, where various missions from \textit{Uhuru}
\citep{uhuru} to \rosat\ have systematically surveyed the sky and
where slew surveys of later missions have added detail, our knowledge
of the sky at hard X-rays ($>10$~keV) has been rather patchy and
insensitive. The sensitivity of the HEAO-A4 13--180~keV survey
\citep{levine84}\ was such that only 77 sources were
detected. 

Recently \integral-IBIS has provided some observations
\citep{3rdBird,beckmann06,beckmann09,krivonos07} that are much more
sensitive but have concentrated on certain regions of the sky; the
exposure in the latest IBIS `all-sky' catalog varies from one part of
the sky to another by a factor of a thousand, some regions of the sky
having only a few thousand seconds of observation.  The RXTE all-sky
slew survey \citep{rxte-slew} covers much of the sky in the 3-20 keV
band and detects 294 sources, but the coverage is not uniform or
complete and the sensitivity is weighted to lower energies such that
the BAT and RXTE sources are not the same.

A survey in the hard X-ray band is important for several
reasons. Observations below 15~keV can be drastically affected by
photoelectric absorption in certain sources, giving a false indication
of their luminosity. Populations of heavily absorbed or Compton-thick
Active Galactic Nuclei (AGN) have been hypothesized in order to
explain the portion of the spectrum of the diffuse hard X-ray
background ascribed to unresolved sources \citep{gilli07}, but such
objects have not been found in the necessary numbers, prompting
questions as to the composition and evolution of a population of AGN
that could explain its form \citep{treister09}. Hard X-ray emission is
also being discovered from an unexpectedly large number of previously
unknown Galactic sources, notably from certain cataclysmic variables,
symbiotic stars and heavily obscured high mass X-ray binaries
\citep{3rdBird}.
 
The Burst Alert Telescope (BAT) on \swift\ \citep{gehrels04} has a
large field of view and is pointed at a large number of different
directions which are well distributed over the sky. The resultant
survey provides the most uniform hard X-ray survey to date and
achieves a sensitivity sufficient to detect very large numbers of
sources, both Galactic and extragalactic. \cite{markwardt05} have
published the results from the first three months of BAT data, and
\cite{tueller9} have published a survey of sources seen in the first 9
months of \swift\ observations, concentrating on the 103 AGN seen at
Galactic latitudes greater than 15\degree. We present here a catalog
of all sources detected in the first 22 months of operations, (2005
Dec 15 -- 2006 Oct 27) increasing the number of AGN to 266 and including
all other sources seen across the entire sky.
  
%--------------------------------------------------------------------
\section{\swiftbat}

\swift\ is primarily a mission for the study of gamma-ray
bursts. \swift\ combines a wide field instrument, BAT, to detect and
locate gamma-ray bursts (GRBs) with two narrow field instruments to
study the afterglows (the X-ray Telescope (XRT) \citep{burrows05} and
the Ultra-Violet/Optical Telescope (UVOT)
\citep{roming05}). \swiftbat\ is a wide field ($\sim$2~sr) coded
aperture instrument with the largest CdZnTe detector array ever
fabricated (5243~cm$^2$ consisting of 32,768 4mm detectors on a 4.2mm
pitch) \citep{barthelmy05}. BAT uses a mask constructed of 52,000
5x5x1~mm lead tiles distributed in a half-filled random pattern and
mounted in a plane 1m above the detector array.

This configuration results in a large field of view and a
point-spread-function (PSF) that varies between 22\arcmin\ in the
center of the field of view (FOV) and $\sim$14\arcmin\ in the corners
of the FOV (50\degree off axis).  When many snapshot images (a
snapshot is the image constructed from a single survey observation of
$\sim5$minutes) are mosaicked together the effective PSF is
$\sim$19.5\arcmin.

Point sources are found using a fast Fourier Transform convolution of
the mask pattern with the array of detector rates; this effectively
uses the shadow of the mask cast by a source onto the detector array
to create a sky image.

Over much of the BAT field of view, the mask shadow does not cover the
whole array.  The partial coding fraction is defined as the fraction
of the array that is used to make the image in a particular direction
and varies across the FOV.  The BAT field of view is 0.34, 1.18 and
2.29~sr for areas on the sky with greater than $95$\%, $50$\% and
$5$\% partial coding fractions.

\swift\ is in low Earth orbit, but because it can slew rapidly it can
avoid looking at the Earth. The narrow field instruments cannot be
pointed within 45\degree of the Sun, within 30\degree of the Earth
limb, or within 20\degree of the Moon.  

\begin{figure*}
\begin{center}
\resizebox{0.8\textwidth}{!}{
%  \rotatebox{0}{\includegraphics{expo_map_3.png}}}
  \rotatebox{0}{\includegraphics{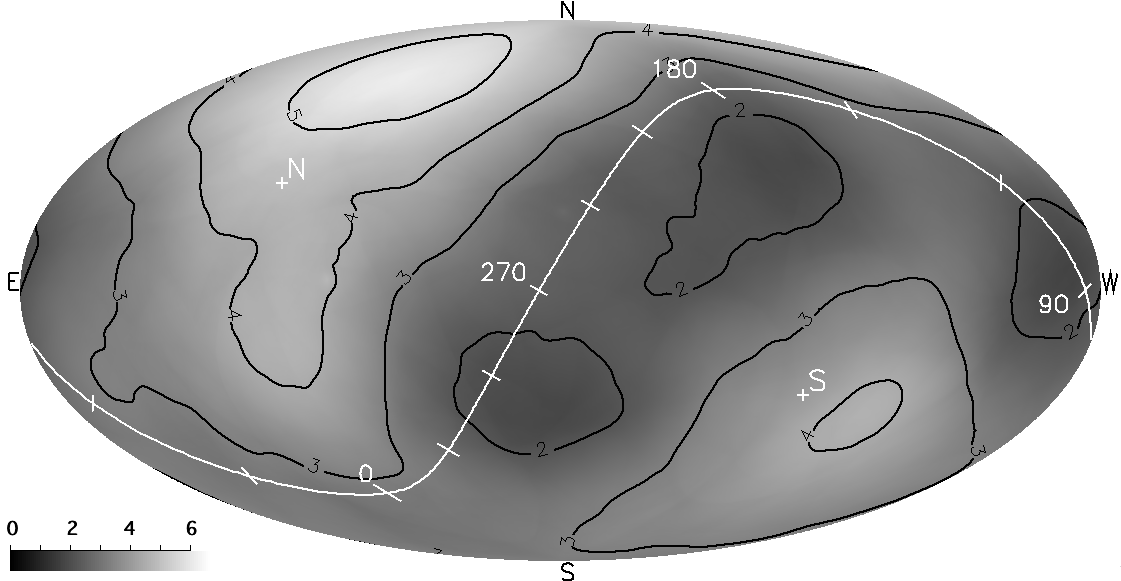}}}
\resizebox{0.8\textwidth}{!}{
%  \rotatebox{0}{\includegraphics{sens_map_4.png}}}
  \rotatebox{0}{\includegraphics{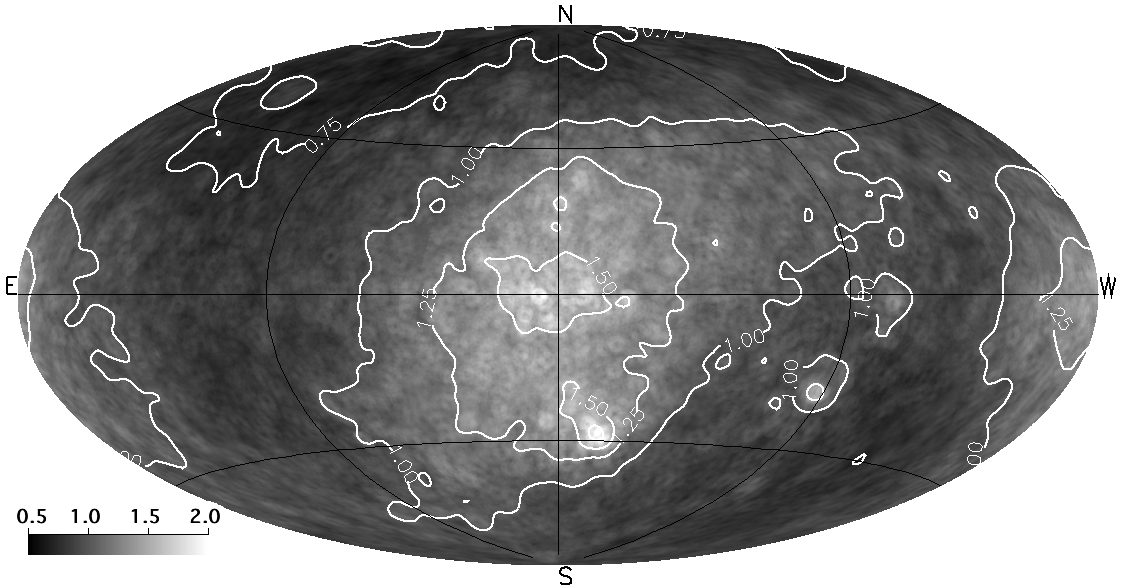}}}
\caption{The top panel shows the effective exposure map for the
  22-month \swiftbat\ survey in a Hammer-Aitoff projection on Galactic
  coordinates.  The ecliptic poles and equator are also shown; the
  largest exposures are toward the north and south ecliptic poles (the
  units on the colorbar are Ms).  The bottom panel shows the measured
  $5\sigma$ sensitivity across the sky in units of mCrab in the
  14--195~keV band. The bright spots at l=344.1, b=-44.0 (GRB060614),
  l=271.8, b=-27.2 (GRB060729), and l=254.7, b=-1.4 (GRB060428a) are
  areas of high systematic noise due to very long exposures
  ($>800$~ks) performed early in the mission before dithering in roll
  angle was instituted.
\label{exposure-map}}
\end{center}
\end{figure*}

The pointing plan for \swift\ is optimized to observe GRBs.  This
strategy produces observations spread out over a few days and at
nearly random positions in the sky.  The BAT FOV is so large that most
of the sky is accessible to BAT on any given day, but the pointing is
deliberately biased toward the anti-Sun direction in order to
facilitate ground based optical follow-up observations of gamma-ray
bursts.  Even though the \swift\ pointing plan is optimized for GRB
observations, BAT's large FOV and \swift's random observing strategy
result in very good sky coverage (50--80\% at $>$20~mCrab in one day)
for transients (\cite{krimm06}\footnote{online at:
  \url{http://swift.gsfc.nasa.gov/docs/swift/results/transients/}}). Over
a longer term, this observing strategy produces an even more uniform
sky coverage (see Figure~\ref{exposure-map}), with an enhanced
exposure at the ecliptic poles caused by avoiding the Sun and Moon.
This high coverage factor means that the BAT survey can provide
reasonably well sampled light curves and average fluxes compiled from
data taken throughout the period covered by the survey.

\begin{figure*}[htb]
\begin{center}
%\resizebox{0.48\textwidth}{!}{\includegraphics{sky_cover_diff_3.pdf}}
\resizebox{0.48\textwidth}{!}{\includegraphics{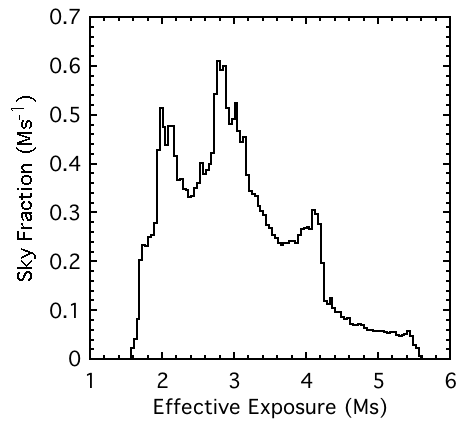}}
\resizebox{0.48\textwidth}{!}{\includegraphics{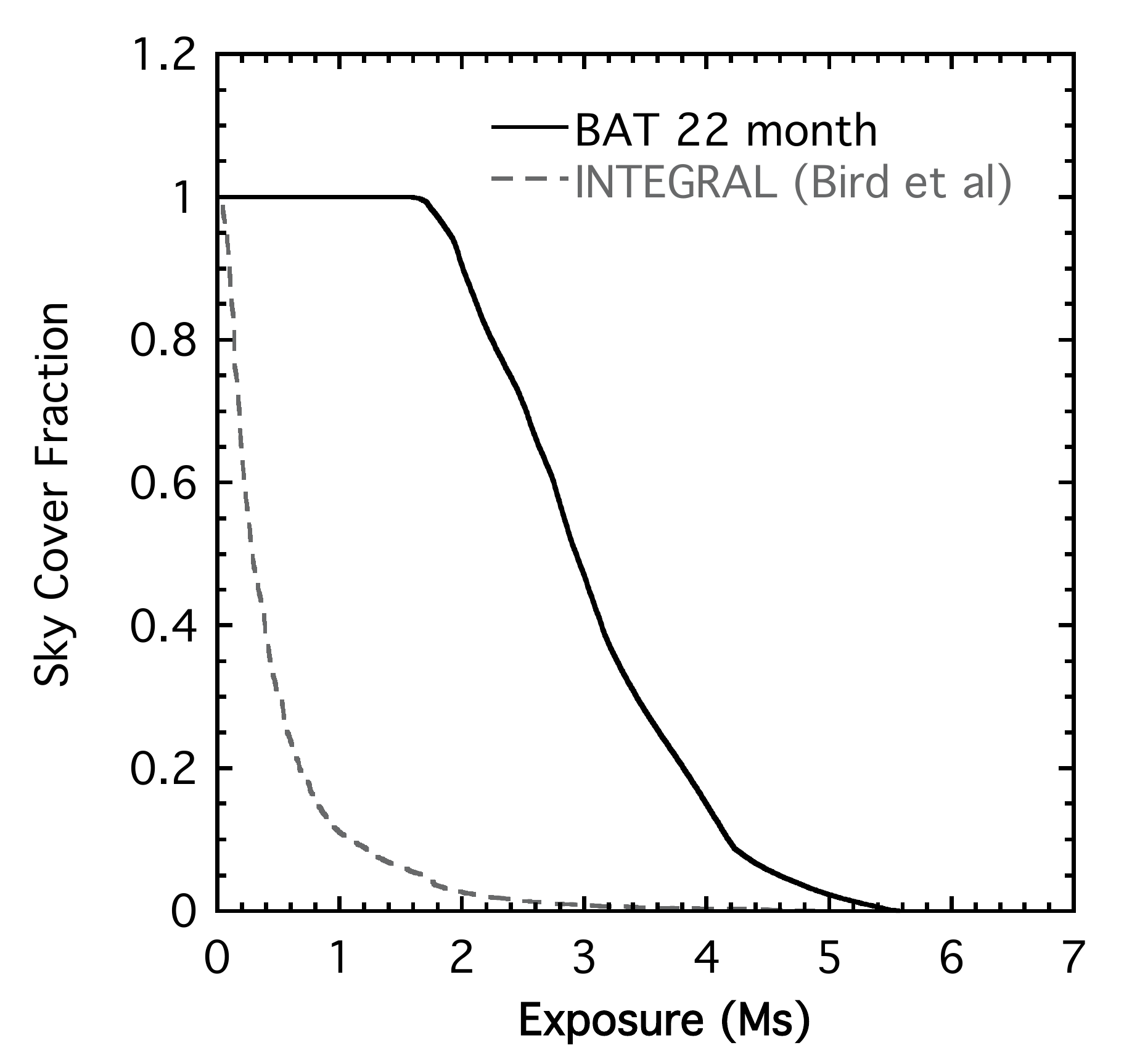}}
\caption{Spatial uniformity in the 22-month BAT hard X-ray survey. The
  left panel is the effective exposure time histogram (bin size =
  40~ks), and the right panel compares the fraction of sky seen by the
  BAT and \integral\ surveys as a function of effective exposure
  times.\label{sky_coverage}}
\end{center}
\end{figure*}

%\subsection{BAT instrument}

The effective exposure time of each point in the field of view is the
equivalent on-axis time (partial coding fraction times observing
time); therefore the observing time for a place on the sky is
generally much larger than is displayed in the effective exposure
map. All of the source count rates from the BAT survey are normalized
to this effective exposure.  

The observing efficiency of \swiftbat\ is high for a satellite in low
Earth orbit, but observing inefficiencies (passages through the South
Atlantic Anomaly (SAA) 16\%, slewing 16\%, down time $<1$\%) still
result in a loss of 33\% of the total observing time. 

The BAT 22-month survey includes data taken between 15 Dec 2004 and 27
Oct 2006; there are 39.6~Ms of usable data in the 22 month survey. A
typical point on the sky is within the BAT FWHM FOV $\sim10$\% of the
time, and the survey data screening rejects 0.5\% of the data (see
\S\ref{filtering}). These two effects result in an effective exposure
of 3.9~Ms for a typical point on the sky.  The histogram of the
effective exposure (Figure~\ref{sky_coverage}) shows that most of the
sky has an effective exposure time between 1.7 and 4.3~Ms, with a few
regions receiving as much as 5.6~Ms.

The BAT PSF is determined by the mask tile cell and detector cell
sizes.  For an on-axis source, the PSF is approximately Gaussian with
a FWHM of 22.5 arcminutes.  In the native Cartesian tangent plane
coordinate system of BAT images, the PSF has nearly a constant
shape and size throughout the field of view.  However, because tangent
plane units are not spaced at equal celestial angles, the true PSF
shape is compressed for off-axis sources, varying approximately as
$\sigma_{\rm PSF} = 22.5^\prime / (1 + \mathrm{tan}^2\theta)$ where $\theta$ is
the angle of the source from the pointing axis.  When averaged over
many pointings, and weighted by partial coding and solid angle, the
mean PSF is $\sim$19.5 arcminutes FWHM.  We use this 19.5 arcminute
PSF when analyzing the BAT survey mosaicked skymaps since they are
composed of many contributing snapshot observations.

%\subsection{Coded mask imaging}

%--------------------------------------------------------------------
\section{BAT Survey Processing}

The following sections describe the BAT survey analysis techniques as
implemented in the \texttt{batsurvey} software tool.  General
information on coded mask imaging can be found in
\cite{skinner,skinner08}, \cite{fenimore}, and \cite{caroli}.

\subsection{BAT Survey Data Collection and Initial Filtering}
\label{filtering}

The BAT instrument monitors the sky in ``survey" mode when not within
a few minutes of responding to a gamma-ray burst. In this mode
detected events are binned into histograms by the instrument flight
software and the histogram counts are periodically telemetered to the
ground (typically on a 5 minute interval).  These histograms contain
detector (spatial) and pulse height (energy) information.  On the
ground the histograms are further adjusted to place all detectors on
the same energy scale, and then for the standard survey analysis are
re-binned into the eight survey energy bands: 14--20~keV, 20--24~keV,
24--35~keV, 35--50~keV, 50--75~keV, 75--100~keV, 100--150~keV, and
150--195~keV.

Several quality filters are applied to the BAT survey data.  First,
the spacecraft must be in stable pointing mode, which means that the
attitude control ``10 arcmin settled" flag must be set.  The
spacecraft star tracker must be reporting ``OK" status, and the
boresight direction must be at least 30$^\circ$ above the Earth's
limb.  Second, BAT must be producing good quality data, which means
that the overall array event rate must not be too high or low
(3000~cts~s$^{-1}$ $<$ rate $<$ 12,000~cts~s$^{-1}$); a count rate
lower than 3000~cts~s$^{-1}$ means that the detector is not operating
correctly, and a rate higher than 12,000~cts~s$^{-1}$ only occurs
during passages through the SAA.  A minimum number of detectors must
be enabled ($>$ 18,000 detectors out of 32,768), and no histogram bins
can be reported as missing data because of bad telemetry.  In
addition, histogram time intervals that cross the UTC midnight
boundary are discarded since the spacecraft has at times been
commanded to make small maneuvers during that time.  These temporal
filters produce a set of good time intervals over which the histograms
are summed.  The finest time sampling of this survey analysis is
approximately a single pointed snapshot (which have durations of
$\sim$150--2000~s).  The good time intervals are further checked so
that the spacecraft pointing does not change appreciably during the
interval (1.5 arcmin in pointing, 5 arcmin in roll), and data are
excluded if the pointing has varied.  Short intervals of 150 seconds
or less are discarded in order to ensure enough counts across the
detector for the balancing stage (see \S\ref{balancing}) of the
processing to work correctly.

After temporal filtering each pointed snapshot is reduced to a set of
eight detector count maps, one for each energy band.  Since the
systematic noise in the sky images depends on the quality of
individual detectors, significant effort is made in optimizing the
spatial filtering of the data (i.e, the masking of undesirable
detectors).  All detectors disabled by the BAT flight software are
masked.  In addition the detector counts maps are searched for noisy
(``hot") detectors using the \texttt{bathotpix} algorithm; any
detectors found to be noisy are masked.  Finally, detectors with known
noisy properties (i.e. high variance compared to Poisson statistics)
are discarded.  The ``fixed pattern'' noise (see \S\ref{patternmap})
is also subtracted from each map.

\subsection{Removal of Bright Sources}
\label{balancing}

Bright point sources and the diffuse sky background contribute
systematic pattern noise to the entire sky image, at approximately 1\%
of the source amplitude, due to the coded mask deconvolution
technique.  By subtracting the contributions of these sources from the
detector images, the systematic noise can be significantly reduced.
We used the \texttt{batclean} algorithm to remove bright sources and
diffuse background at the snapshot level.  The diffuse background is
represented as a smooth polynomial in detector coordinates.  Bright
sources are represented by the point source response in the detector
plane.  The source responses are generated using ray tracing to
determine the shadow patterns.  Bright sources are identified by
making a trial sky map, and any point source detected above $9\sigma$
in any energy band is marked for cleaning.  In our experience--and
based on the properties of the BAT mask---no new bright sources become
detectable after the \texttt{batclean} calculation, so it is not
necessary to iterate the process again.  In order to preserve the
original bright source intensities, we insert the fluxes from the
uncleaned maps into the cleaned maps around the locations of these
sources.

At the \texttt{batclean} stage the maps are also ``balanced" so that
systematic count rate offsets between large scale spatial regions on
the detector are removed.  Sources shining through the mask do not
produce this kind of coherent structure, therefore this balancing
stage helps to remove systematic noise.  This process involves
dividing the array into detector module sides (128 detectors,
[$=16\times 8$]), which are separated by gaps of 8.4--12.6~mm from
neighboring detector module sides.  The mean counts in both the outer
edge detectors (44 detectors), and the inner detectors (84 detectors),
are subtracted for each module separately, so that the mean rate is as
close as possible to zero.  Count rate variations from module to
module are believed to occur because of variations in the quality of
CZT detector material and because of dead time variations in the
module electronics caused by noisy pixels.  Variations between outer
edge and inner detectors in each module are due to cosmic ray
scattering and X-ray illumination of detector sides.  The BAT coded
mask modulates the count rate of cosmic sources on essentially
detector-to-detector spatial scales, so the subtraction of the mean
count rates averaged over many tens of detectors does not affect the
coded signal.

Very bright sources which are partially coded will cast shadows of the
mask support structures on the edges of the mask.  These shadows are
not coded by the mask, are highly energy dependent, and thus must be
treated carefully.  This is done by masking detectors in regions of
the detector plane affected by mask-edge regions for bright sources
($\sim$0.3~Crab or brighter) determined via ray tracing.

After subtracting bright sources and background, detectors whose
counts are more than $4\sigma$ from the mean are discarded in order to
further remove contributions from noisy detectors.

\subsection{Fixed Pattern Noise}
\label{patternmap}

Non-uniform detector properties cause variations between the
background count rates measured in different detectors.  These spatial
differences form a relatively stable pattern over timescales much
longer than a day and are not addressed by the \texttt{batclean}
algorithm.  These spatial differences also comprise a fixed noise in
detector coordinates which is transformed by the survey processing
into unstructured noise in the sky image.  This fixed pattern is
determined by constructing long term averages of the residual BAT
count rates of each individual detector, after subtracting the
contributions of bright sources as described above. In this
construction, variable terms average to zero and only the stable
pattern remains.  This pattern maps are then subtracted from each
snapshot detector image.  The benefit of removing this fixed pattern
noise is that each individual detector is addressed, thereby removing
systematic noise on a finer scale than the balancing stage mentioned
previously.

The contributions of some detectors to this fixed pattern is time
dependent as a result of temporal variations in detector
performance. We address this time dependence in each detector by
fitting a polynomial to the daily average value.  The fits are done
on data spanning weeks to many months, and the polynomial used has
$\sim 1$~order per 30 days fit.

This approach (subtracting the long-term average
fixed pattern noise from the data) avoids removing any legitimate
signal from sources since \swift\ changes its pointing direction on
much shorter time scales.

In practice, the entire survey processing must be run once initially
for all of the data, with the pattern contribution set to zero, in
order to determine the residual rates mentioned above.  Once the
pattern maps have been computed, the processing is run a second time
using those values.

\subsection{Sky Maps}

Sky maps are produced for each snapshot using the \texttt{batfftimage}
algorithm which cross-correlates the detector count maps with the mask
aperture pattern.  Sky maps are sampled at 8.6~arcmin on-axis, which
corresponds to half the natural element spacing for the coded mask.
The natural sky projection for these maps is tangent-plane; thus, the
sky-projected grid spacing becomes finer by a factor of $\sim$2 at the
extreme edges of the field of view.  The angular extent of a sky map
from one snapshot covers the region in the sky where the BAT has some
non-zero response.  This field of view is approximately $120^\circ
\times 60^\circ$, although the sensitivity is much reduced at the
edges of the field of view due to projection effects through the mask
(foreshortening of the mask and shadowing due to the mask thickness at
large off-axis angles) and partial coding.

The snapshot maps are corrected for partial coding, geometric
projection effects, and the number of active detectors.  Thus,
they represent the BAT count rate per fully illuminated detector,
corrected approximately to the on-axis response.  An examination of
the measured count rates of the Crab nebula (considered to be a stable
point source for the BAT) shows some systematic residual trends as a
function of off-axis angle and energy.  These effects are primarily
due to absorption by passive materials in the field of view, whose
absorption lengths scale approximately as $\mathrm{sec}\,\theta$, where
$\theta$ is the off-axis angle.  The absorptions can be as high as
50\% at the lowest energies and largest angles, but are typically
smaller.  After correction for these effects, the count rate estimates
are accurate to within a few percent.

Partial coding and noise maps which represent the partial exposure of
each pixel in the sky map are created for each pointed snapshot.  The
partial coding maps are further adjusted to correct for the fact that
some parts of the sky are occulted by the Earth during the
observation.  For each observation, a map of the average Earth
occultation is computed showing the fraction of the observation time
each pixel is occulted, and the partial coding map is multiplied by
this occultation map to account for the reduced effective observing
time.  The noise maps are generated by computing the local r.m.s.\ of
the pixel values in an annulus around each position (see
\S\ref{blind_search}).

Individual pointed snapshot sky maps are discarded if the differences
between the model used in the bright source removal and the cleaned,
binned detector plane data lead to a reduced chi-square value greater
than 1.25.  This filtering excluded primarily data around the bright
X-ray source Sco~X-1, which produces such strong count rate
modulations that they aren't reduced to zero at the Poisson
statistical level by \texttt{batclean}.  This does produce a
significant exposure deficit around Sco~X-1 and the Galactic center
region.

\subsection{Mosaicking}

The sky images from each snapshot are weighted by inverse variance
(i.e. noise$^{-2}$) and combined into all-sky maps.  Each snapshot sky
map contributing to the mosaic is trimmed such that all areas of the
snapshot have greater than 15\% partial coding.  The sky is divided
into six facets in Galactic coordinates, with grid spacing of the
pixels 5 arcmin at the center of each facet.  The Zenithal Equal Area
projection was used in order to minimize distortion far from the
center of projection.  Each individual sky image is projected and
resampled onto the all-sky grids by bilinear interpolation, as are the
partial coding and noise maps.  The final result is a set of weighted
flux maps, propagated noise maps and effective exposure maps for each
energy band and facet combination, plus an additional one for the
total energy band of 14--195 keV.

This analysis procedure produces a sky image where each pixel
represents the best estimate of the flux for a point source at the
corresponding position in the sky (see \cite{fenimore} for more
information on coded mask image reconstruction).

\subsection{Source Detection}
\label{blind_search}

A ``blind" source detection algorithm was used to search for sources
in the mosaicked significance maps using the full survey bandpass of
14--195~keV.  The significance map is the ratio of the counts map to
the local noise map.  

The RMS noise map is calculated from the mosaicked sky map using an
annulus of radius 30~pixels (2.5\degree) with an inner exclusion
radius of 8 pixels (40\arcmin).  An 8~pixel radius around the position
of all known BAT sources is also excluded from the regions used for
background calculation.  We do not attempt to fit the PSF of the
source for the noise calculation and hence the positions of sources
must be eliminated from this calculation to get an accurate measure of
the underlying noise in the image.

The noise is assumed to be a smooth function of image position and so
the value at the center of the annulus is well approximated by the
average value in the annulus. This calculated noise includes both
statistical and systematic noise and is therefore a better estimate of
the total noise in the image than the noise calculated from a PSF
fit. The noise from every source is distributed over the whole image,
just as the signal from the source is distributed over the detector
array, so no local enhancement of noise at the position of the source
is expected.

The blind search algorithm first finds all peaks in the map by
searching for pixels that are higher than each of the surrounding 8
pixels.  If the significance in a peak pixel is greater than our
detection threshold of $4.8\sigma$ (see \S\ref{significance}), the
excess is considered to be a detection in the blind search for new
sources.

%--------------------------------------------------------------------
\section{The \swiftbat\ 22-Month Catalog}
\label{catalog}

\begin{deluxetable}{rlr}
%\tabletypesize{\small}
\tablecaption{Counterpart Types in the \swiftbat\ 22-month Catalog\label{type-decomp}}
\tablewidth{\columnwidth}
\tablehead{
\colhead{Class} & \colhead{Source Type} &
\colhead{\# in catalog}
}
\startdata
0       & Unidentified\tablenotemark{a}         & 19 \\
1       & Galactic \tablenotemark{b}           & 3 \\      
2       & Extragalactic\tablenotemark{c}        & 17 \\
3       & Galaxy Clusters      & 7 \\
4       & Seyfert Galaxies                  & 229 \\
5       & Beamed AGN\tablenotemark{d}           & 32 \\
6       & CVs / Stars          & 36 \\
7       & Pulsars / SNR        & 15 \\
8       & X-ray Binaries       & 121
\enddata
%\tablerefs{(1) \citealt{angr89}; (2) \citealt{grsa98}.}
%\tablecomments{Source class of the 476 counterparts in the
%  \swiftbat\ 22-month catalog.}
\tablenotetext{a}{Sources listed as unidentified have an object with
  unknown physical type as a counterpart.  Some of these objects are
  associated with a source detected at another wavelength.}
\tablenotetext{b}{Sources classified as galactic are so assigned
  because of observed transient behavior in the X-ray band along with
  insufficient evidence to place them in another class.}
\tablenotetext{c}{Sources in the extragalactic class are seen as
  extended in optical or near-IR imagery, but do not have firm
  evidence (such as an optical spectrum) from other wavebands
  confirming whether they harbor an AGN.}  \tablenotetext{d}{Sources
  classified as ``beamed AGN'' include blazars, BL~Lacs, FSRQs,
  quasars, and other high redshift AGN.}
\end{deluxetable}

The catalog of sources detected by \swiftbat\ using the first 22
months of data includes sources at all Galactic latitudes.  The
22-month catalog and associated data in electronic form can be found
online at the
\swift\ website.\footnote{\texttt{http://swift.gsfc.nasa.gov/docs/swift/results/bs22mon/}}

\begin{figure*}
\begin{center}
\resizebox{0.9\textwidth}{!}{\includegraphics{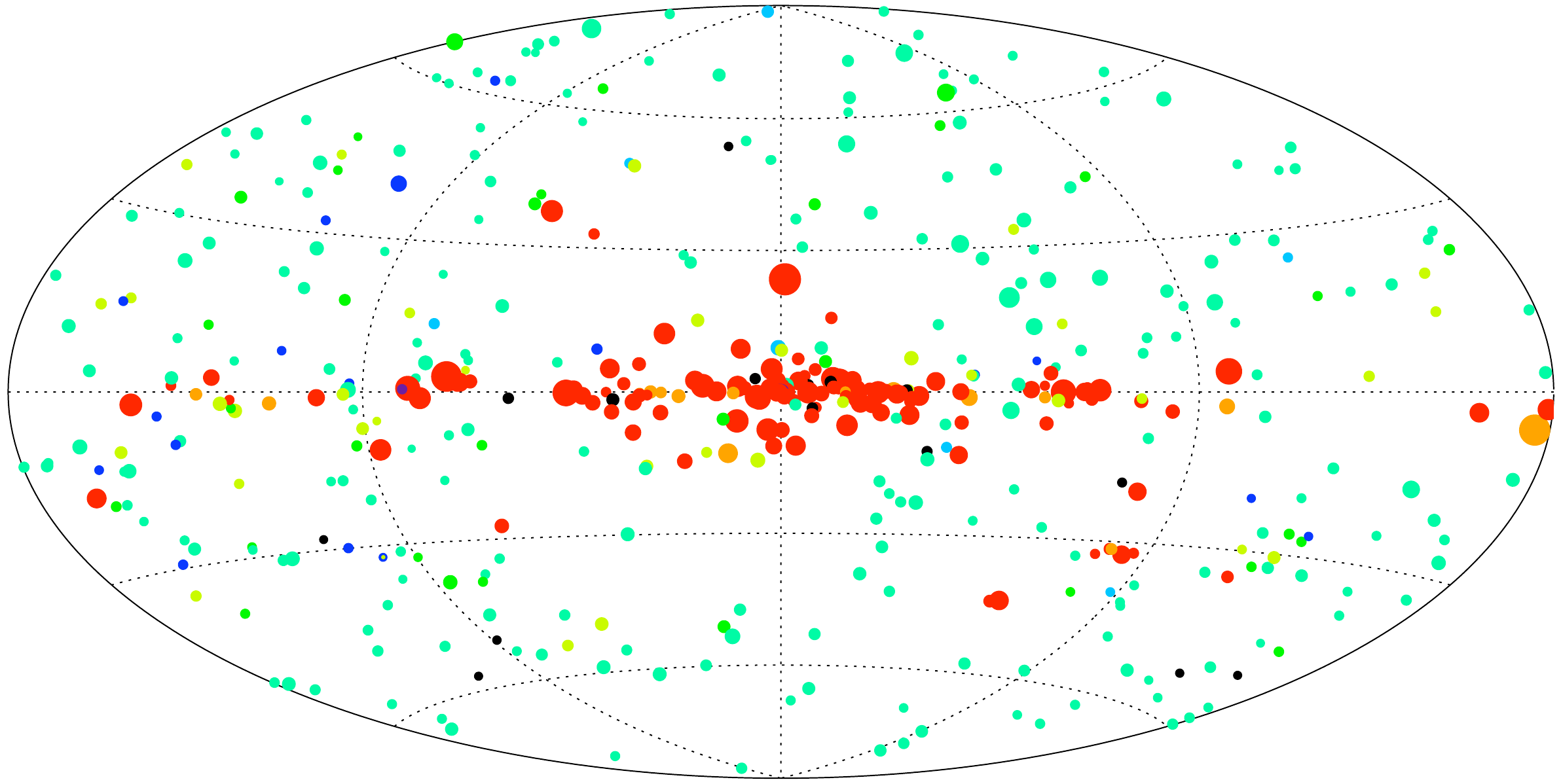}}
\resizebox{0.9\textwidth}{!}{\includegraphics{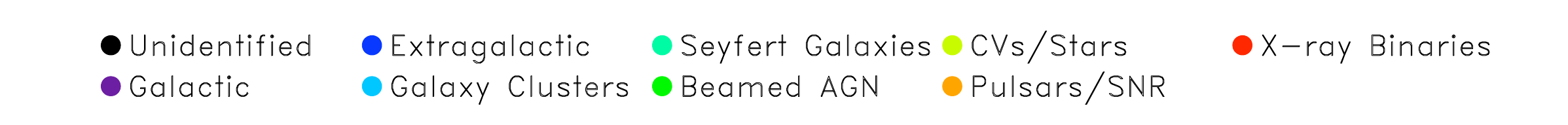}}
\caption{All sky map showing classification of the BAT 22-month survey
  sources.  The figure uses a Hammer-Aitoff projection in Galactic
  coordinates; the flux of the source is proportional to the size of
  the circle.  The source type is encoded by the color of the
  circle.\label{aitoff-type}}
\end{center}
\end{figure*}

Figure~\ref{aitoff-type} shows the distribution of sources on the sky
color coded by source type, with the symbol size proportional to the
source flux in the 14--195~keV band.  Table~\ref{type-decomp} gives
the distribution of objects according to their source type.  Sources
classified as ``unidentified'' are those where the physical type of
the underlying object (e.g., AGN, CV, XRB, etc) is unknown.  These
sources have a primary name derived from the BAT position.  Some
unidentified BAT sources are associated with sources in the X-ray or
gamma-ray bands (with positions unable to sufficiently determine an
optical counterpart or physical source type), and these sources can be
distinguished by having a name in the catalog derived from the
observation in the other waveband.  The few sources classified only as
``Galactic'' generally lie in the plane and have shown some transient
behavior which indicates a Galactic source, but no other information
is available that would allow further
classification. ``Extragalactic'' sources are detected as extended
sources in optical or near-IR imaging, but do not have other
indications of being an AGN.  The ``Beamed AGN'' category includes
BL~Lacs, blazars, and FSRQs.

Table~\ref{table_sources} is the listing of all the sources detected
above the 4.8$\sigma$ level in a blind search of the 22-month
\swiftbat\ survey maps.  The first column is the source number in the
22-month catalog.  The second column of the table is the BAT name,
constructed from the BAT source position given in columns three and
four.  In cases where the source has been previously published with a
BAT name corresponding to a slightly different location (e.g., a source
position from a previous BAT catalog with less data), we have used the
first published name but have given the correct 22-month BAT
coordinates in columns two and three.  The fifth column is the
significance of the blind BAT source detection in sigma units.
Instances where more than one possible counterpart to a
single BAT source is likely are indicated with ditto marks in columns 2--5.

The sixth column gives the name of the identified counterpart to the
BAT hard X-ray source with the most precisely known position.  These
are often optical galaxies, or 2MASS sources, and are associated with
a source detected in the medium-energy X-ray band (3--10~keV) in
\chandra, \xmm, or XRT images.  Counterpart determination is discussed
in \S\ref{counterparts}.  The seventh column gives an alternate name
for the counterpart.  We have preferred to list a well known name (e.g.,
Sco~X-1) or a name from a hard X-ray instrument or high energy
detection.  The best available coordinates of the counterpart (J2000)
are given in the table in columns~8 and 9.

The 10th and 11th columns give the 14--195~keV flux of the BAT source
(in units of 10$^{-11}$~ergs~sec$^{-1}$~cm$^{-2}$) and its $1\sigma$
error.  The BAT flux for each counterpart is extracted from the hard
X-ray map at the location of the counterpart given in columns 8 and 9.
The flux determination method is described in \S\ref{flux}.

The 12th column indicates whether there is source confusion: there is
source confusion either if there is more than one possible XRT
counterpart or if two likely hard X-ray sources lie close enough
together to make a proper extraction of the flux not possible with the
standard method.  The treatment of confused sources is discussed in
more detail in \S\ref{confused}.  We define two classes of source
confusion: ``confused'' sources, and ``confusing'' sources. A source
is ``confusing'' for the purposes of this column if a fit to the map
indicates that the source contributes to the hard X-ray flux of a
neighboring source.  A ``confused'' source has received more than 2\%
of its flux from a neighboring source.  A confused source is labeled
with an ``A'' in this column, and a confusing source with a ``B'' (the
case of a very bright source next to a weak one would result in the
bright source labeled with a ``B'' and the weak source with an ``A'').
A source that is both confused and confusing (e.g., the case where there
are two similar strength sources close to each other, such as when
there are two possible XRT counterparts to a single BAT source) is
labeled with an ``AB''.

When a source has an entry in column 12, a best estimate of the
counterpart flux is listed in column 10 from a simultaneous fit of all
the counterparts in the region to the BAT map. When the entry is ``A''
or ``AB'' in column 12 (indicating a confused source), the error on
the flux is not well defined, and column 11 is left blank. (See
\S\ref{flux}).

The 13th and 14th columns list the source hard X-ray hardness ratio
and its error computed as described in \S\ref{hardness}.  The hardness
ratio is defined here as the ratio of the count rate in the 35--150~keV
band divided by the count rate in the 14--150~keV band.

The 15th and 16th columns give the redshift and BAT luminosity of the
counterpart if it is associated with a galaxy or AGN.  The source
luminosity (with units log[ergs~s$^{-1}$] in the 14--195~keV band) is
computed using the redshift and flux listed in the table and a
cosmology where $H_0 = 70$~km~s$^{-1}$~Mpc$^{-1}$, $\Omega_m = 0.30$,
and $\Omega_\Lambda = 0.70$.

The 17th column lists a source type with a short verbal description of
the counterpart.

%The ***18th and last column gives a numerical source
%class as follows: 0 = unidentified, 1 = galactic, 2 = extragalactic
%(associated with a galaxy), 3 = cluster of galaxies, 4 = AGN, 5 =
%blazar, 6 = CV or star, 7 = pulsar or SNR, and 8 = X-ray binary.

\subsection{Source Positions and Uncertainties}

The BAT position is determined by using the BAT public software tool
\batcelldetect\ to fit the peak in the map to the BAT PSF (a two
dimensional Gaussian with a FWHM of 19.5~arcminutes). The
\batcelldetect\ program performs a least-square fit using the local
rms noise to weight the pixels in the input map.  These fit positions
were used to generate the BAT positions in the catalog and the names
of newly detected BAT sources.  

The PSF fit using \batcelldetect\ also reports a formal position
uncertainty based on the least-square covariance matrix.  However,
because neighboring pixels in the coded mask images are inherently
correlated, the formal uncertainty reported by this technique will not
be representative of the true uncertainty.  Therefore we choose to use
the offset between the fit position and the counterpart position as an
indicator of the BAT position error.

The \batcelldetect\ program also has the option of fitting source
locations using an input catalog of starting positions.  We have used
this capability to test the stability of the source positions found by
\batcelldetect\ by using an input catalog where all the starting
positions have been offset by 8 arcminutes in a random direction from
the source position found in the blind search. We have performed this
test with several different offsets and find that the fit converges to
within 1 arcminute of the counterpart position for BAT sources that
are not confused.  For a few sources, the fit sometimes converges onto
a side peak instead of the primary peak, but this error is not
repeated in additional tests starting from other randomized positions.
This type of systematic error in the position determination does not
occur in the blind search (\S\ref{blind_search}) since we use the
maximum pixel to start the fit instead of a randomized spot
8~arcminutes from the blind position.  Anomalous offsets in the source
position are identified by examination of the image and refitting.

\begin{figure}
\begin{center}
\resizebox{\columnwidth}{!}{\includegraphics{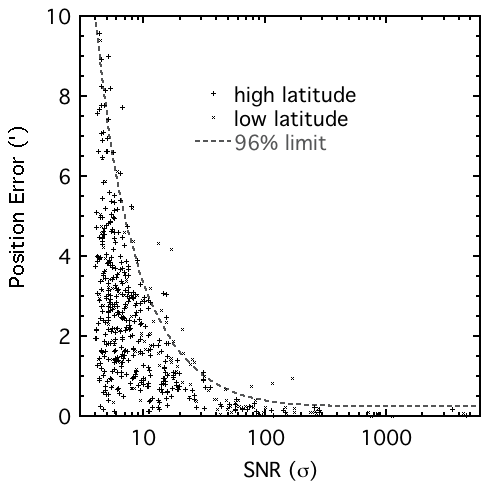}}
\caption{The BAT position error as a function of the BAT detection
  significance. The angular separation between the counterpart
  position and the fitted BAT position is used to determine a measured
  position error for each source.  This measured position error is
  plotted as a function of BAT detection significance.  The dashed
  line in the plot shows the 96\% error radius as a function of BAT
  source detection significance.  Sources with large position errors
  are almost always low galactic latitude sources falling in regions
  of high source density and locally higher noise.  \label{ang_sep}}
\end{center}
\end{figure}

In order to judge the accuracy of the BAT positions, we plot in
Figure~\ref{ang_sep} the angular separation between the BAT position
and the counterpart position against the significance of the BAT
source detection.
The accuracy of the BAT position improves as the significance of the
detection becomes stronger.  There are 461 BAT sources in
Table~\ref{table_sources} with detection significances greater than
$4.8\sigma$; there are 479 possible counterparts, and of these 25 are
located grater than 5 arcminutes from the BAT position.  Therefore,
there is only a $\sim5$\% chance of a BAT-detected source ($>
4.8\sigma$) having a counterpart farther away than 5 arcminutes.

In Figure~\ref{ang_sep} we also plot a line showing our estimate of
the BAT position error for a given source significance.  This estimate
for the error radius (in arcminutes) can be represented with the
function
\begin{equation}
\mathrm{BAT~error~radius} = 
\sqrt{\left(\frac{30}{\left(S/N-1\right)}\right)^2 +
  \left(0.25\right)^2},
\end{equation}
where $S/N$ is the BAT detection significance.  This empirical
function includes a systematic error of 0.25~arcmin deduced from the
position errors of very significant sources.  This error radius
includes 96\% of the sources that are greater than $5^\circ$ from the
Galactic plane and $15^\circ$ from the Galactic center.  The error
radius encloses 85\% of sources in the Galactic plane.  Sources known
to be confused are not included in the plot.

\subsection{Counterparts}
\label{counterparts} 

Counterparts to the BAT sources were primarily discovered by examining
X-ray images taken with instruments with good angular resolution.
\chandra\ resolution is sometimes required on the plane, otherwise
\xmm, Suzaku or \asca\ images were examined.  \rosat\ images and source
catalogs were of relatively low importance for counterpart
identification because of \rosat's lack of effective area in the hard
X-ray band, because of the poor correlation between \rosat\ flux and
the BAT hard X-ray flux (see \cite{tueller9}, Figure~7), and because
of the high chance probability of finding a \rosat\ source in the BAT
error circle.

If no archival X-ray images existed for the location of a BAT source,
we requested \swift-XRT followup observations of the field containing
the BAT source.  A 10~ks observation with XRT is deep enough to detect
almost all BAT sources.  BAT extragalactic sources are usually AGN
contained in bright ($J\sim 13$), nearby galaxies at redshift $z <
0.1$ and are easily identified in an XRT observation.  

The X-ray counterpart to an unabsorbed BAT source is a very bright XRT
source, which is easily detected with a 2~ks XRT observation. However,
most of the new BAT sources are heavily absorbed in the X-ray band and
were not detected by ROSAT. We have found empirically that XRT can
detect essentially all of the BAT sources (including the absorbed
ones) in a 10 ks observation.

We require consistency of the BAT and the X-ray spectrum ($>3$~keV)
when simultaneously fit with an absorbed power law allowing only a
renormalization between BAT and XRT to account for variability. This
consistency of the spectra is required for all sources not previously
known to be hard X-ray emitters, except transients and sources known
to have highly variable spectra where the BAT spectrum averaged over
years cannot be directly compared to the XRT measurement from a single
observation.

A small fraction of XRT follow-up observations in the 5--10~ks range
detected multiple sources consistent with the BAT position.  In these
cases the counterpart to the BAT source was almost always identified
by limiting the bandpass of the X-ray image to the higher energy
3--10~keV band.  This bandpass filtering usually reduced the number of
sources in the field to a single hard source.

In the few cases where two or more hard sources still remain after
bandpass filtering, all are considered possible counterparts to the
BAT source and listed in the catalog with a flag indicating that the
counterpart identification suffers from source confusion. There are 18
more possible counterparts in Table~\ref{table_sources} than there are
blind BAT sources (461) because of the 15 cases where there are one
or more possible counterparts to a single BAT source.

Because the counterpart identification requires an X-ray point source
with a small error radius ($\sim 4$~arcsec), a positional coincidence
with a known source or bright galaxy, \textbf{and} an X-ray spectrum
consistent with the BAT flux, we believe that the counterpart
misidentification rate is extremely small.  All of the identified
counterparts listed in Table~\ref{table_sources} are hard X-ray
sources.

\subsection{Confused Sources}
\label{confused}

Sources are labeled as confused in our table when the highest pixel
associated with the BAT source in the mosaicked maps (the ``central
pixel'' value) has a significant contribution from adjacent sources.
This includes the cases when two possible X-ray counterparts lie
within a single BAT pixel and when two BAT sources are close enough
that each contributes flux to the location of the adjacent source.

Using the positions of the X-ray counterparts as an input catalog, we
calculated the fractional contribution of each BAT source to its
neighbors. We used the significance measured in the blind search and
the 19.5 arcminute FWHM Gaussian BAT PSF to calculate the intensity of
each source at the position of its neighbors.  The central pixel value
for each BAT source was assumed to be the sum of the source plus all
the contributions from its neighbors.  This creates a set of linear
equations that can be solved for the true significance of each source.
We solved these equations with the constraint that sources were not
allowed to have negative significance.  This procedure was devised to
determine cases where the BAT significance is altered because of the
presence of a very strong nearby source.

If we found that the resulting fit S/N from the technique that
accounted for contributions from neighbors was lower than the central
pixel S/N from the blind search by 2\%, we labeled it as confused.

\subsection{Detection Significances and Limits}
\label{significance}

The detection significance for the BAT sources in the catalog is
extracted from the mosaicked significance map at the BAT position (see
\S\ref{blind_search}).  The significance is taken from the highest
pixel value in the blind search.

\begin{figure}
\begin{center}
\resizebox{\columnwidth}{!}{\includegraphics{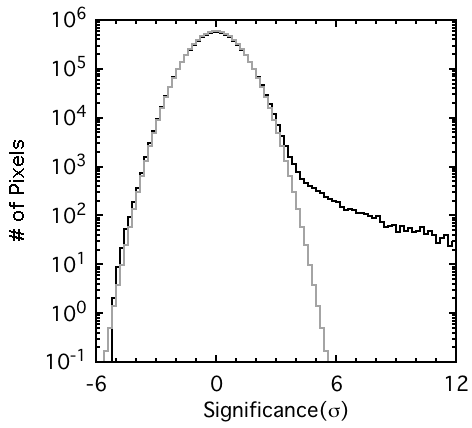}}
\caption{Histogram showing the significances of the pixels in the
  22-month survey.  The gray line is not a fit to the data; it is a
  Gaussian distribution with $\sigma = 1$ and normalized to the peak
  of the observed distribution.
\label{gaussian_pix}}
\end{center}
\end{figure}

Figure~\ref{gaussian_pix} shows the distribution of individual pixel
significances from the mosaicked map of the entire sky.  
As is usual for a coded mask imager, the noise distribution is a
Gaussian function centered at zero significance, with a width of
$\sigma = 1$ and a total integrated area equal to the number of pixels
in the map.  The large tail at positive significance is due to real
astrophysical sources present in the map.

The distribution of the pixel significances in
Figure~\ref{gaussian_pix} closely follows a Gaussian distribution for
the negative significances.  The positive side of the distribution
also follows a Gaussian, but with the addition of pixels with enhanced
significances because of the presence of real sources in the map.

Examination of the negative fluctuations provides a good measure of
the underlying noise distribution.  There is only 1 negative
pixel in the entire map with a magnitude greater than
$5\sigma$.
We therefore choose our detection limit to be $4.8\sigma$.  This
detection limit is also the same as used in the 9-month version of the
\swiftbat\ catalog.  While it is clear that there are several real
sources with significances somewhat smaller than $4.8\sigma$, we
choose this value in order to minimize false sources caused by random
fluctuations.  We expect random fluctuations to account for 1.54
sources at the $4.8\sigma$ level in our sky map of $1.99\times 10^6$
independent pixels.

\begin{figure}
\begin{center}
\resizebox{\columnwidth}{!}{\includegraphics{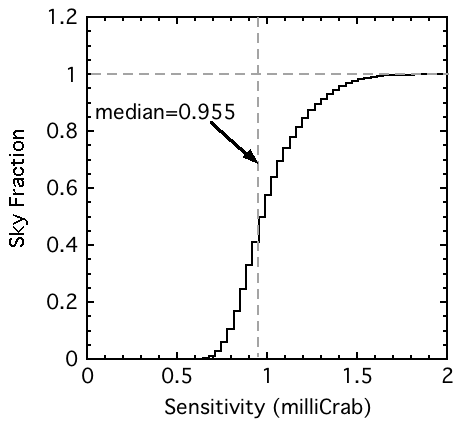}}
\caption{The integral distribution of sky coverage versus sensitivity
  achieved in the survey.  The 1~mCrab sensitivity limit (for 50\% sky
  coverage) corresponds to a flux of $2.3 \times
  10^{-11}$\,ergs\,cm$^{-2}$\,s$^{-1}$ in the 14--195~keV band.
\label{sens_sky}}
\end{center}
\end{figure}

Figure~\ref{sens_sky} shows the integral distribution of sky coverage
versus sensitivity achieved in the survey.
We achieve a sensitivity of better than 1~mCrab for half the sky, which
corresponds to a flux of $< 2.3 \times
10^{-11}$\,ergs\,cm$^{-2}$\,s$^{-1}$ in the 14--195~keV band.

\subsection{Counterpart Fluxes}
\label{flux}

Fluxes of the counterparts to BAT sources were extracted from the
mosaicked maps using the pixel containing the position of the
identified counterpart.  For sources where a counterpart is not known,
we use the fitted BAT position to determine the flux.

We have chosen to normalize source fluxes in the 8 survey bands to the
Crab because the systematic uncertainties in the survey averaged Crab
spectrum are smaller than the uncertainties in the BAT survey response
matrix.  The source fluxes in each band were computed by comparing
the source count rate to the measured rate of the Crab Nebula in each
band:
\begin{equation}
\mathrm{BAT\;source\;flux = \left(\frac{BAT\;source\;count\;rate}
{Crab\;count\;rate}\right) Crab\;flux },
\end{equation}
where the Crab flux in each band is given by
\begin{equation}
\mathrm{Crab}\;\mathrm{flux} = \int_{a}^{b} 
E\;F(E)\:dE ,
\end{equation}
where $a$ and $b$ are the lower and upper BAT band edges and $E$ the
energy in keV.   

We take the Crab counts spectrum to be
\begin{equation}
F(E) = 10.17\:E^{-2.15} 
\left(\frac{\rm photons}{{\rm cm}^2\,{\rm sec}\:{\rm keV}}\right),
\end{equation}
determined by fitting a power-law model to BAT on-axis calibration
observations taken early in the \swift\ mission.  These values are
consistent with characterizations of the Crab spectrum using data from
Integral/SPI \citep{integral/spi}, Integral/IBIS
\citep{integral/ibis}, HETE/FREGATE \citep{hete}, SAX/PDS
\citep{sax/pds}, and GRIS \citep{gris}.

The total Crab flux is then
\begin{equation}
\mathrm{Crab}\;\mathrm{flux} = \int_{14\;keV}^{195\;keV} 
E\;F(E)\:dE = 2.44\times 10^{-8}
\left(\frac{\rm ergs}{{\rm cm}^2\,{\rm sec}}\right).
\end{equation}

Sources with a spectral index very different from the Crab can have a
small but significant residual systematic error in the fluxes
determined with this method.

In order to gauge this error we generated counts spectra for different
models in the eight survey bands using the BAT on-axis spectral
response matrix.  The Crab comparison flux determination method
described above was used to obtain the model fluxes in each of the
eight survey bands.  The flux errors between the computed fluxes and
the model fluxes in the individual bands were always $<10$\% for a range
of model spectral indices between 1 and 3 and so we deem this
technique acceptable for producing the source fluxes in each of the
survey bands.

We fit the 8-channel spectra with a power law model in order to
produce an overall hard X-ray flux for each BAT source.  We used
\texttt{XSPEC} and a diagonal matrix to fit the 8-channel spectra with
the \texttt{pegpwrlw} model over the entire 14--195~keV BAT survey
energy range in order to extract the source flux in this band.  This
approach was selected because it weights the energy bands by their
individual uncertainties; a simple sum of the bands would produce a
very large error due to the high weight it assigns to the noisiest
bands at the highest energies. 

The $1\sigma$ error in the overall flux was determined by using the
error function in XSPEC and is given in Table~\ref{table_sources}.
For the highest significance BAT sources ($>100$~sigma), this
procedure does not produce a good fit (reduced $\chi^2 \gg 1$), but
this is to be expected from the very high significances of each point
and the coarse energy binning. To evaluate the systematic error in the
fitting we performed fits to our model spectra generated from the
response matrix. For power law spectra, the systematic error in the
flux is dominated by the error in the individual data points as
calculated above.  Sources with hardness ratios less than 0.1 are not
well fit with a power law, and the systematic uncertainty in the flux
can be significantly larger.

The fluxes for sources marked as confused were calculated in a
slightly different way.  Instead of using the count rate extracted
from the map at the counterpart position, we performed a simultaneous
fit to find the fluxes of all the sources in the confused region as
described in \S\ref{confused}. For these sources we do not quote an
error on the flux estimate because the behavior of the errors with
this fitting technique is not well known.  Any source with a confused
flag should be considered as detected by BAT but the flux should be
considered as an upper limit.

\subsection{Sensitivity and Systematic Errors}
\label{syserr}

In this section, we compare the expected statistical errors with the
actual measured statistical noise in the final mosaic maps.  From the
perspective of pure Poisson counting statistics, the uncertainties are
governed primarily by the properties of the coded mask and the
background (see \cite{skinner08} for details).  The expected
$5\sigma$ noise level can be expressed as (adapting from
\cite{skinner08} Eqn.~23 and 25):
\begin{equation}
5\sigma_{\rm Poisson} = 5 \sqrt{\frac{2b}{\alpha\:N_{\rm det}\:T}},
\label{eq:poinoise}
\end{equation}
where $b$ is the per-detector rate, including background and point
sources in the field of view; $N_{\rm det}$ is the number of active
detectors ($N_{\rm det} \le 32768$); $T$ is the effective on-axis
exposure time; and $\alpha$ is a coefficient dependent on the mask
pattern and detector pixel size ($\alpha = 0.733$ for BAT).  The
partial coding, $p$, enters the expression through the ``effective
on-axis exposure'' time, $T = p T_o$, where $T_o$ is the actual
exposure time.  Using nominal values
($b=0.262$~cts~s$^{-1}$~detector$^{-1}$; $N_{\rm det} = 23500$ (the
exposure-weighted mean number of enabled detectors); and
Crab rate = $4.59\times 10^{-2}$~cts~s$^{-1}$~detector$^{-1}$), we find
the estimated Poisson $5\sigma$ noise flux level to be
\begin{equation}
\label{eq:poiflux}
f_{5\sigma} = 0.99\:{\rm mCrab} \left(\frac{T}{1\:{\rm Ms}}\right)^{-1/2}.  
\end{equation}
We consider this to be a lower limit to the expected Poisson noise
level for a given effective exposure.  In reality, the background rate
$b$ may be higher than the nominal value by up to 50\% depending on
the particle environment of the spacecraft.  Also, along the Galactic
plane the contributions of bright sources such as the Crab, Sco~X-1
and Cyg~X-1, are not strictly negligible, and will raise the overall
level of $b$ by up to $\sim$10\%.  All of these adjustments would
cause a Poisson noise level larger than given by equation
\ref{eq:poiflux}, by an amount that depends on the specific satellite
conditions during the survey.  We estimate that, averaged over the
entire survey duration, the true Poisson noise level may be 5--15\%
higher than the lower limit quoted above.

\begin{figure}
\begin{center}
\resizebox{\columnwidth}{!}{\includegraphics{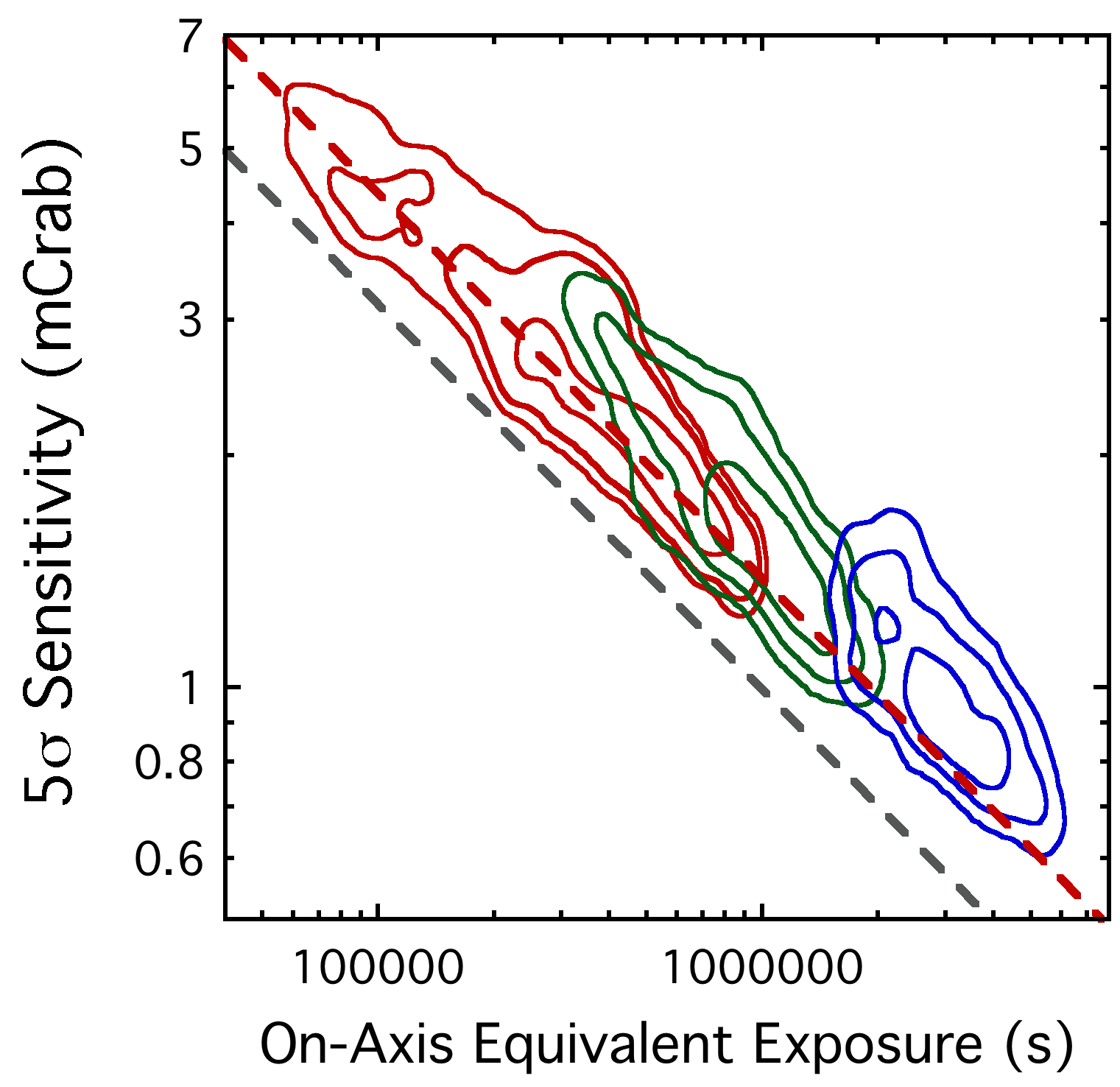}}
\caption{Measured $5\sigma$ BAT sensitivity limit for pixels in the
  all-sky map, as a function of effective exposure time, $T$, for the
  3-month (red; \cite{markwardt05}), 9-month (green; \cite{tueller9})
  and 22-month (blue; this work) survey analyses.  The contours
  indicate the number of pixels with a given sensitivity and effective
  exposure.  The contour levels are linearly spaced.  The red dashed
  line represents the original $T^{-1/2}$ sensitivity curve quoted in
  \cite{markwardt05}.  The black dashed line represents a lower limit
  to the expected Poisson noise level (see \S\ref{syserr}).  The
  measured noise is approximately 30--45\% higher than the expected
  Poisson noise.
  \label{sens_expo} }
\end{center}
\end{figure}

Figure~\ref{sens_expo} compares the measured noise and expected noise
versus effective on-axis exposure.  
We see that both noise measures are decreasing approximately as
$T^{-1/2}$, which suggests that the dominant errors are uncorrelated
over time.
It also suggests that pointing
strategies such as roll-angle dithering have been successful in
reducing pointing-related systematic errors.  However, the measured
noise is still higher than the expected Poisson noise by
$\sim$30--45\%, and we take this to be a measure of the unmodeled
systematic variations on the detector plane.

The largest likely contributors to systematic variations are improper
subtractions of diffuse background, and of bright sources.  BAT count
rates are background-dominated --- the background rate is equivalent
to $\sim$6--9 Crab units --- so the coded mask analysis is
particularly sensitive to imperfect subtraction of spatial background
variations.  While the pattern map method and the functions fitted
during the cleaning stage produce a good model of the detector
background, some imperfections remain.  One effect is that
detector-to-detector sensitivity differences, coupled with varying
exposures to the X-ray background, can lead to excess residuals.

For similar reasons, bright sources may also contribute systematic
noise.  The brightest sources are clustered along the Galactic plane,
and thus contribute noise in those preferred locations.  Indeed, we
note that the measured noise is $\sim$50\% higher in the Galactic
center region, where there is a concentration of bright point sources.
This is a larger factor than can be accounted for by a larger count
rate.  Modeling of point sources may be imperfect for the same reasons
as for the background.  Also, there may be other effects such as side
illumination of detectors that may contribute additional noise.  Here,
``side illumination'' refers to the facets of the individual CdZnTe
detectors which do not face the pointing direction, but are still
sensitive to X-rays.  Off-axis sources will shine through the mask and
illuminate the sides, producing an additional (although fainter) coded
signal.  At the moment, illumination of the sides of detectors is not
modeled at the imaging or cleaning steps, and thus there will be an
additional noise due to the effect.

Proper modeling of these systematic error contributors will be the
subject of future work.  At this stage we do not have an in-depth
analysis of the quantitative contributions of each effect to the
systematic noise, and in some cases the analysis may be prohibitively
difficult.  Still, at the current exposure levels, the noise level
seems to be decreasing with exposure, and we do not appear to be
reaching an ultimate systematic limit in this analysis.

\subsection{Spectral Analysis}
\label{hardness}

In \S\ref{flux} we use a simple power-law fit to the data to estimate
the source flux.  However, because the catalog contains sources with
various different physical natures and spectral shapes, we choose to
use the more robust characterization given by a hardness ratio to
describe the BAT spectra.

\begin{figure*}
\begin{center}
\resizebox{0.9\textwidth}{!}{
  \rotatebox{0}{\includegraphics{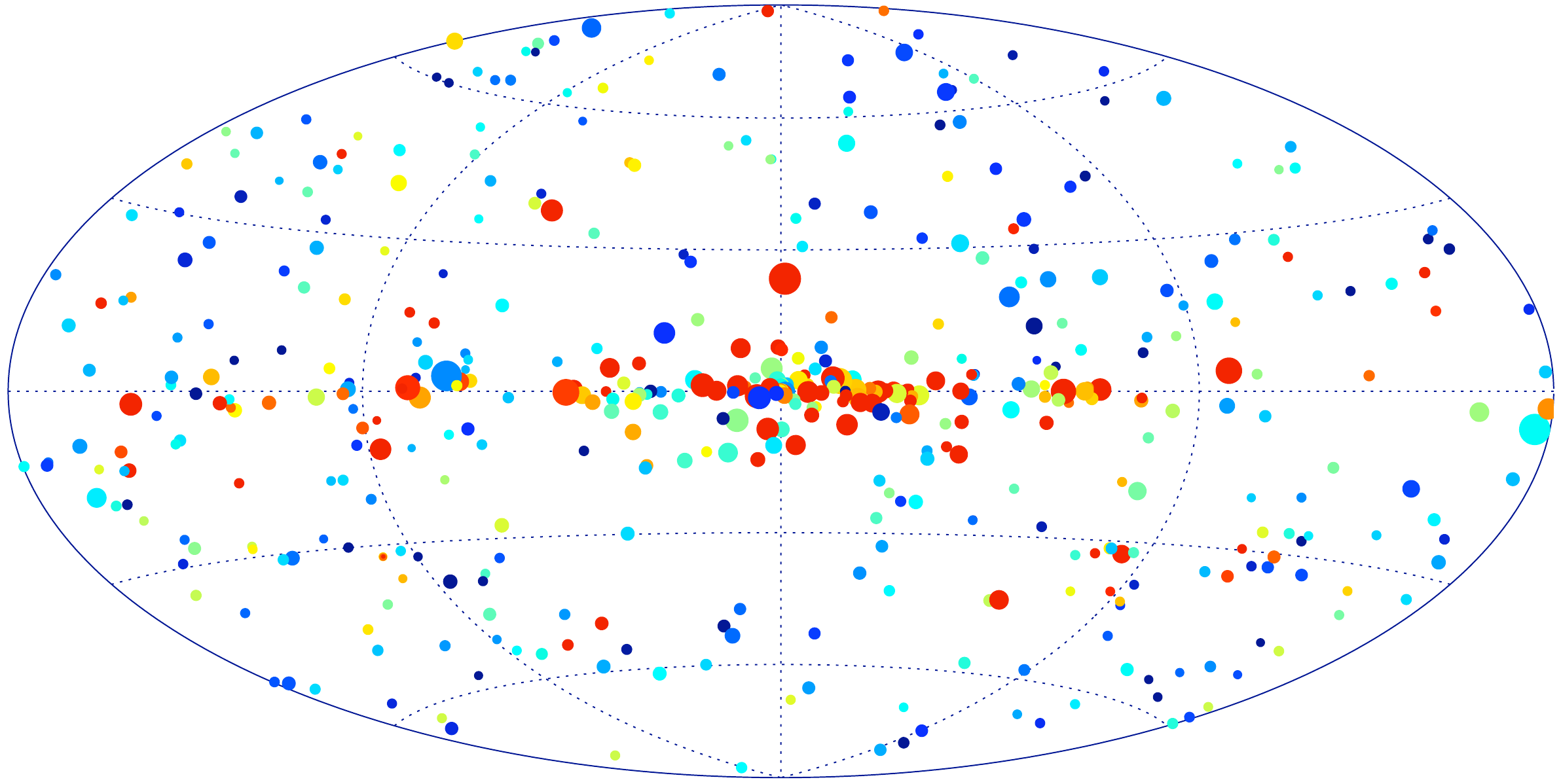}}}
\caption{All sky map showing spectral hardness of BAT 22-month survey
  sources.  The figure uses a Hammer-Aitoff projection in Galactic
  coordinates; the size of the circle is proportional to the flux of
  the source.  Blue sources are harder, and red sources are
  softer.\label{aitoff-hardness}}
\end{center}
\end{figure*}

\begin{figure*}
\begin{center}
\resizebox{0.98\textwidth}{!}{\includegraphics{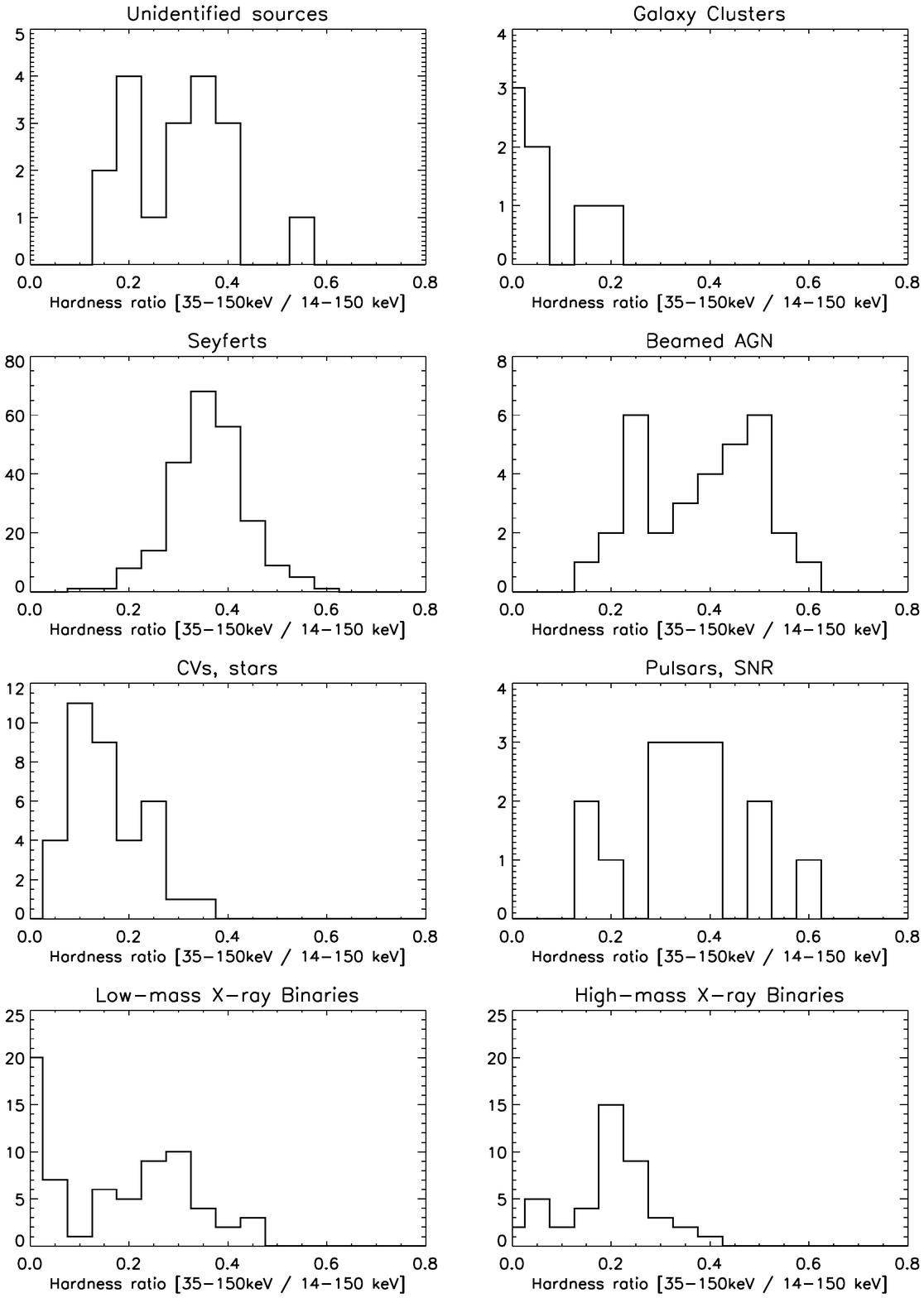}}
\caption{Hardness ratios of BAT 22-month sources by source class.
%  (*** fill in) 
\label{hardness_ratios}}
\end{center}
\end{figure*}

Hardness ratios for the \swiftbat\ sources were calculated by taking
the sum of the count rate in the 35--150~keV band and dividing by the
count rate in the 14--150~keV band.  Errors on the hardness ratio were
calculated by propagating the errors on the count rates in the
individual 8 bands except when the source is listed as confused.
Figure~\ref{aitoff-hardness} shows a map of the source positions on
the sky, with the source flux represented by the size of the point and
the source hardness by the color (red is softer and blue is harder).
Figure~\ref{hardness_ratios} shows the hardness ratios of the 22-month
BAT sources by source class.

\begin{figure}
\begin{center}
\resizebox{\columnwidth}{!}{
  \rotatebox{0}{\includegraphics{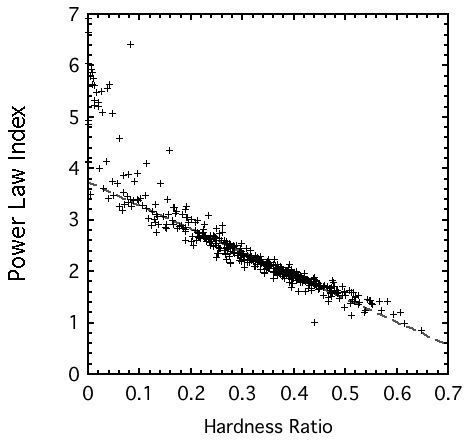}}}
\caption{The correlation between power law index and hardness ratio
  for the BAT survey sources.\label{gamma_to_pwrlaw}}
\end{center}
\end{figure}

A mapping can be made between hardness ratio and power-law index for
sources that have spectral shapes well described by a simple power-law
model (e.g., the majority of AGN in our catalog).  This mapping is
well represented by 
\begin{equation}
\Gamma = 3.73 - 4.52\ HR,
\end{equation}
where $\Gamma$ is the power-law index and $HR$ is the hardness ratio
as defined above.  Figure~\ref{gamma_to_pwrlaw} shows the correlation
between power-law index and hardness ratio for the BAT survey sources.
The correlation holds well for sources with hardnesses above about
0.15, but begins to break down for softer sources.  An illustration of
this problem is the soft BAT spectra of clusters of galaxies, which
have thermal spectra with temperatures $\sim10$~keV and are usually
detected only in the lowest energy BAT band and are not well fit by a
power-law model.  We leave to a later paper a more careful spectral
analysis using models appropriate to the physical nature of the
sources.

%----------------------------------------------------------------------
\section{\swiftbat\ Survey Sources}

Although it is for extragalactic astronomy that the present survey
represents the greatest step forward, a comparison of the results for
Galactic sources with earlier work also has interesting implications.

Of the 479 sources in Table~\ref{table_sources}, 97\% have reasonably
firm associations either with objects known in other wavebands or with
previously known X-ray or gamma-ray sources.  More than 60\% of the
associations are with extragalactic objects. At high Galactic
latitudes ($|b|>10^\circ$) the density of identified extragalactic
sources is 22.6~sr$^{-1}$, and it is only only slightly reduced at low
latitudes to 19.2~sr$^{-1}$. This suggests that only $\sim 7$
extragalactic sources in the plane are missed through reduced
sensitivity, lack of information in other wavebands, or confusion, and
illustrates the uniformity of the survey.

153 BAT AGN were previously reported in the BAT 9-month AGN survey
\citep{tueller9}.  \cite{winter09} provide X-ray spectral fits for
these sources and provide measures of the luminosity, \nh\ , etc.\ for
the sources in the 9-month catalog.  Most of the sources in the
9-month catalog also appear here in the 22-month catalog; however,
variability in the sources has caused 11 sources to drop out of the
22-month list that were in the 9-month catalog.

%%%
\subsection{New sources}

\tabletypesize{\small}
\begin{deluxetable*}{p{38mm} l p{98mm}}
%\begin{center}
%\rotate
\setlength{\tabcolsep}{5pt}
\tablecaption{New high energy  sources in the 22-month catalog that are Galactic, or probably Galactic,
  and were first detected as hard X-ray sources
  by \swiftbat\ \label{new_galactic}}
\tablewidth{0pt}
\tablehead{\colhead{Source} & \colhead{First} & \colhead{Notes} \\  & \colhead{reported} & }
\startdata
SWIFT J0026.1+0508 & Here & Of several sources in an XRT follow-up
                              observation, only one is hard and it is
                              taken to be the counterpart.  It could
                              be a CV.\\

% Gone from catalog
%SWIFT J0030.0$-$5904         & Here       & A faint XRT source in the BAT error circle coincides with 
 %                      a UV-bright star USNO-A2.0 0300$-$00143700\\
%
% Gone from catalog  - BUT WHY ?
% 1RXS J045707.4+452751      & Here       & XRT and UVOT provide precise localization of ROSAT source. 
 %                       Nothing known about this source.\\
%
% Gone from catalog
%SWIFT J052522.48+241331.8  & Atel 1279  & CV. Transient discovered by BAT at ROSAT position 
  %                         XRT-refined position gives identification with USNO-A2.0 1125$-$02389994.\\
  %
  % Gone from catalog, but WHY ?
%SWIFT J0525.8+4256         & ATel 853   & Discovery announcement notes possible association with old SNR 
 %                  SNR 166.0+04.3  (VRO 42.05.01).  BAT position corresponds with radio hot-spot. \\

SWIFT J0732.5$-$1331         & ATel 697   & CV of subtype DQ~Her. ATels 757, 760, 763.\\
%
% Gone from catalog BUT WHY?
% 1RXS J074616.8$-$161127 & Here & First detection as a hard X-ray source   
%	              (??? check basis of association) \\ 
SWIFT J1010.1-5747 & ATel 684 & = CD$-$57 3057.  ATel 669 gives XRT position for BAT source
                         SWIFT J1011.1-5748 = IGRJ 10109-5746
                         associated with Symbiotic star CD-57 3057
                         (ATel 715).\\
%%%%% This source is a galaxy; should not be in this table of galactic
%%%%% sources.
%SWIFT J1453.3+2558         & Here       & There is only one bright source in the XRT field.
%                          It corresponds to RX J1453.1+2554. Galaxy in \cite{bev98}.
%                         Extended in 2MASS.\\
SWIFT J1515.2+1223 & Here & In a 7400~s XRT follow-up, only one source
                          is detected in the hard band at
                          $(\alpha,\delta) =
                          (15\,14\,47,~+12\,22\,44)$.  No known
                          counterparts at this position.\\ 
SWIFT J1546.3+6928 & Here & The BAT source is confused. There are TWO
                          hard ($>$3~keV) sources in the XRT image,
                          1RXS J154534.5+692925 AND 2MASS
                          J15462424+6929102. There are some
                          indications that the ROSAT source is
                          extended, perhaps an interacting pair
                          (making a possible third source).  The 2MASS
                          object is extended and clearly a galaxy.\\ 

SWIFT J1559.6+2554 & ATel 668/9 & = T CrB.  The Swift source is identified with this
                          symbiotic star in XRT follow-up.\\

SWIFT J1626.9$-$5156 & ATel 678 & Peculiar (HMXB?) transient. 15.37~s
                          pulsations.  Optical counterpart
                          2MASS16263652$-$5156305.  Short (100--1000s)
                          flares \citep{reig08}.\\
%% Gone - down to 3.7-4.6 sigma
%SWIFT J1646.2$-$0348         & Here       & XRT field empty in 7.2 ks observation.\\
SWIFT J1753.5$-$0130 & ATel 546 & Short period (3.2~hr; ATel 1130) BH
                          LMXB transient observed with many other
                          instruments following BAT detection.\\

SWIFT J1907.3-2050 & Here & = V1082 Sgr.  XRT follow-up shows a strong hard source coincident
                with the pulsating variable star. \cite{steiner88}
                have found that this star in its hard state has
                properties similar to a CV of subtype
                DQ~Her.  \cite{thorstensen09} has determined an
                orbital period of 20.821~hr for this object which
                classifies it as a long period CV.\\
    
SWIFT J1922.7$-$1717 & ATel 669 & Transient observed with RXTE and
                  Integral after BAT detection \citep{falanga06}\\
%
% Gone - 3.57, 4.82, 4.82 sigma
% SWIFT J1933.?+034? & Here & No XRT follow-up data at time of writing\\
 % \hfill        (UNKNOWN-02)   &        &  within 130" of bright (B=7) emission line variable star V* V1294 Aql  \\
%
SWIFT J1942.8+3220 & Here & = V2491 Cyg.  We find a weak hard X-ray source whose position is
                   consistent with V2491 Cyg in data taken before its
                   eruption as Nova Cyg 2008b.  (see
                   also \cite{ibarra09}, ATel~1478)\\

SWIFT J2037.2+4151 & ATel 853 & Transient; later seen with Integral
                                (ATel 967) \\

SWIFT J231930.4+261517 & ATel 1309 & XRT data show that this source is
                 the same as 1RXS J231930.9+261525, reported and
                 identified as a CV of subtype AM~Her in ATel 1309.
                 Mkn 322 and UGC 12515 may also contribute to the BAT
                 counts\\

SWIFT J2327.6+0629         & Here       & There is no clear source in the XRT field.
\enddata
%\end{center}
\end{deluxetable*}

During the survey BAT has detected a number of new sources that are
transients or other Galactic objects not previously reported as hard
X-ray sources. Some of these have been reported in Astronomer's
Telegrams or elsewhere, others appear for the first time in this
compilation. For convenience these are summarized in
Table~\ref{new_galactic}.  

\tabletypesize{\small}
\begin{deluxetable*}{lrrp{10cm}}
%\rotate
\setlength{\tabcolsep}{5pt}
\tablecaption{Unidentified New Sources\label{new_noident} }
\tablewidth{0pt}
\tablehead{\colhead{Source} & \colhead{$l$\tablenotemark{a}} (\degr) & \colhead{$b$}(\degr)& \colhead{Notes}}
\startdata
%
%SWIFT J0457.1+4528      & 160.97	& 1.52           &  Possible association with 1RXS J045707.4+452751,  but no apparent optical/IR/Radio counterpart. \\
%
SWIFT J0826.2$-$7033   & 284.21	&  $-$18.09  &   = 1ES 0826$-$703,  1RXS J082623.5$-$703142. The  4.1" radius XRT position is 1.2" from T Tau star 2MASS  J08262350$-$7031431. \\
SWIFT J1515.2$+$1223     &   16.44     & +53.28      & Nearest XRT source is a weak one 7.8' away, just outside the 5' radius BAT error circle.\\
SWIFT J1546.3$+$6928  &  104.27  &    40.74    & Two hard XRT sources
lie within the BAT error circle. One is associated with 1RXS
J154534.5+692925, about which nothing is known, the other is
coincident with the extended source 2MASX J15462424+6929102, which is identified with LEDA 2730634, a side-on spiral galaxy.\\
SWIFT J1706.6$-$6146   & 328.72	& $-$12.40    & = IGRJ17062$-$6143. Bright XRT counterpart gives precise position but no apparent optical/IR/Radio counterpart. \\
SWIFT J1709.8$-$3627   & 349.55	& 2.07       & = IGR J17098$-$3628. IGR J17091$-$3624 is only 8.5' away. XRT provides positions for both (ATel 1140). The  BAT position corresponds to  IGR J17098$-$3628, but the XRT error circle contains a complex of IR sources and a radio source and it is not clear which are counterparts.
%  SWIFT J1747.6$-$2817 &     & 0.08 & (we need some work; nearby but distinct XMM, Chandra, IGR sources) \\
%  XTE J1901+014  &    & & Precise XMM position available \\  % Thus is a fast transient
 %  IGR J19443+2117   &57.76	$-$1.30    & ? & XRT shows hard source Maliza et al. (2007) ; Confirms identification with 1RXH J194356.1+211824\\
\enddata
%\vspace{5mm} 
\tablecomments{These sources have new information but no firm identification with an
optical/IR/Radio object. }
\tablenotetext{a}{Galactic coordinates are given as an indication of whether the sources is likely to be Galactic.}
\end{deluxetable*}

In Table~\ref{new_noident} we note other sources that are detected in
this survey and where XRT follow-up has provided additional
information, but where a unique optical, IR, or radio counterpart is
still lacking, or where there is only a BAT detection. Some of these
are almost certainly Galactic objects as may be judged from their
proximity to the Galactic plane.

\begin{deluxetable*}{llccl}
%\tabletypesize{\small}
%\rotate
\tablecaption{New AGN Detected in the \swiftbat\ 22-month Survey with
  Optical Spectroscopic Confirmation\label{new_agn_optical}}
\tablewidth{0pt}
\tablehead{\colhead{BAT Name} & \colhead{Host Galaxy} &
  \colhead{Optical Spectrum\tablenotemark{a}} & \colhead{Galaxy Type} & \colhead{redshift}}
\startdata
SWIFT J0100.9-4750 & 2MASX J01003490-4752033 & 6df  & Sy1.8 & \\       
SWIFT J0623.8-3215 & ESO 426- G 002          & 6df  & Sy2   & \\ 
%SWIFT J0800.1+2324 & CGCG 118-036            & SDSS & Sy2   & \\ 
SWIFT J0923.9-3143 & 2MASX J09235371-3141305 & 6df  & Sy1.8 & \\ 
SWIFT J1513.8-8125 & 2MASX J15144217-8123377 & 6df  & Sy1.8 & \\ 
\\
SWIFT J0249.1+2627 & 2MASX J02485937+2630391 & KP   & XBONG\tablenotemark{b} & 0.058 \\ 
SWIFT J0353.7+3711 & 2MASX J03534246+3714077 & KP   & Sy2   & 0.01828 \\ 
SWIFT J0543.9-2749 & MCG -05-14-012          & KP   & XBONG & 0.0099 \\ 
SWIFT J0544.4+5909 & 2MASX J05442257+5907361 & KP   & Sy1.9 & 0.06597 \\ 
SWIFT J1246.6+5435 & NGC 4686                & KP   & XBONG & 0.0167\\ 
SWIFT J1621.2+8104 & CGCG 367-009            & KP   & Sy2   & 0.0274 \\ 
SWIFT J1830.8+0928 & 2MASX J18305065+0928414 & KP   & Sy2   & 0.019 \\ 
SWIFT J2118.9+3336 & 2MASX J21192912+3332566 & KP   & Sy1   & 0.0507 \\
SWIFT J2341.8+3033 & UGC 12741               & KP   & Sy2   & 0.0174 
%SWIFT J0709.4-1108 & 2MASX J07094208-1107527  \\ 
%SWIFT J1439.2+1417 & 2MASX J14391186+1415215 & \\ 
%SWIFT J1453.1+2556 & 2MASX J14530794+2554327  \\ 
\enddata
\tablenotetext{a}{The optical spectra sources are as follows: 6df = Six degree
  field galaxy survey, SDSS = Sloan Digital Sky Survey, KP = 2.1~m at Kitt Peak.}
\tablenotetext{b}{XBONG = (hard) X-ray Bright, Optically Normal Galaxy}
\end{deluxetable*}

%\begin{deluxetable}{ll}
%\tabletypesize{\small}
%\rotate
%\tablecaption{New AGN Detected in the \swiftbat\ 22-month Survey
%  \label{new_agn}}
%\tablewidth{0pt}
%\tablehead{\colhead{BAT Name} & \colhead{Host Galaxy} } 
%\startdata
%\enddata
%\tablerefs{(1) \citealt{angr89}; (2) \citealt{grsa98}.}
%\tablecomments{Source class of the 476 sources in the
%  \swiftbat\ 22-month catalog.}
%\tablenotetext{a}{These numbers are the photospheric values, used as
%the default in {\tt XSPEC}.}
%\tablenotetext{b}{These numbers are a straight average of the
%photospheric and meteoritic values (except for oxygen, which has the
%updated value given in \citealt{apla01}).}
%\end{deluxetable}

Table~\ref{new_agn_optical} list the new AGN discovered in the
22-month \swiftbat\ survey.  Table~\ref{new_agn_optical} lists those
objects discovered with BAT whose AGN nature could be confirmed with
an optical spectrum.  In column 3 of the table we list the source of
the optical data, and columns 4 and 5 list our own typing of the
spectrum and the redshift.  The optical spectra are mostly obtained
from data in the public domain such as SDSS or 6df, but in a few cases
we have obtained data from our own observations taken at the 2.1m
telescope on Kitt Peak.

%%%
\subsection{Extragalactic Sources}

Most of the extragalactic identifications are with relatively nearby
Seyfert galaxies and many of the remainder are with beamed AGN
(blazars, BL~Lac, FSRQ, etc) sources at much higher $z$.

\begin{figure*}
\begin{center}
\resizebox{\textwidth}{!}{\includegraphics{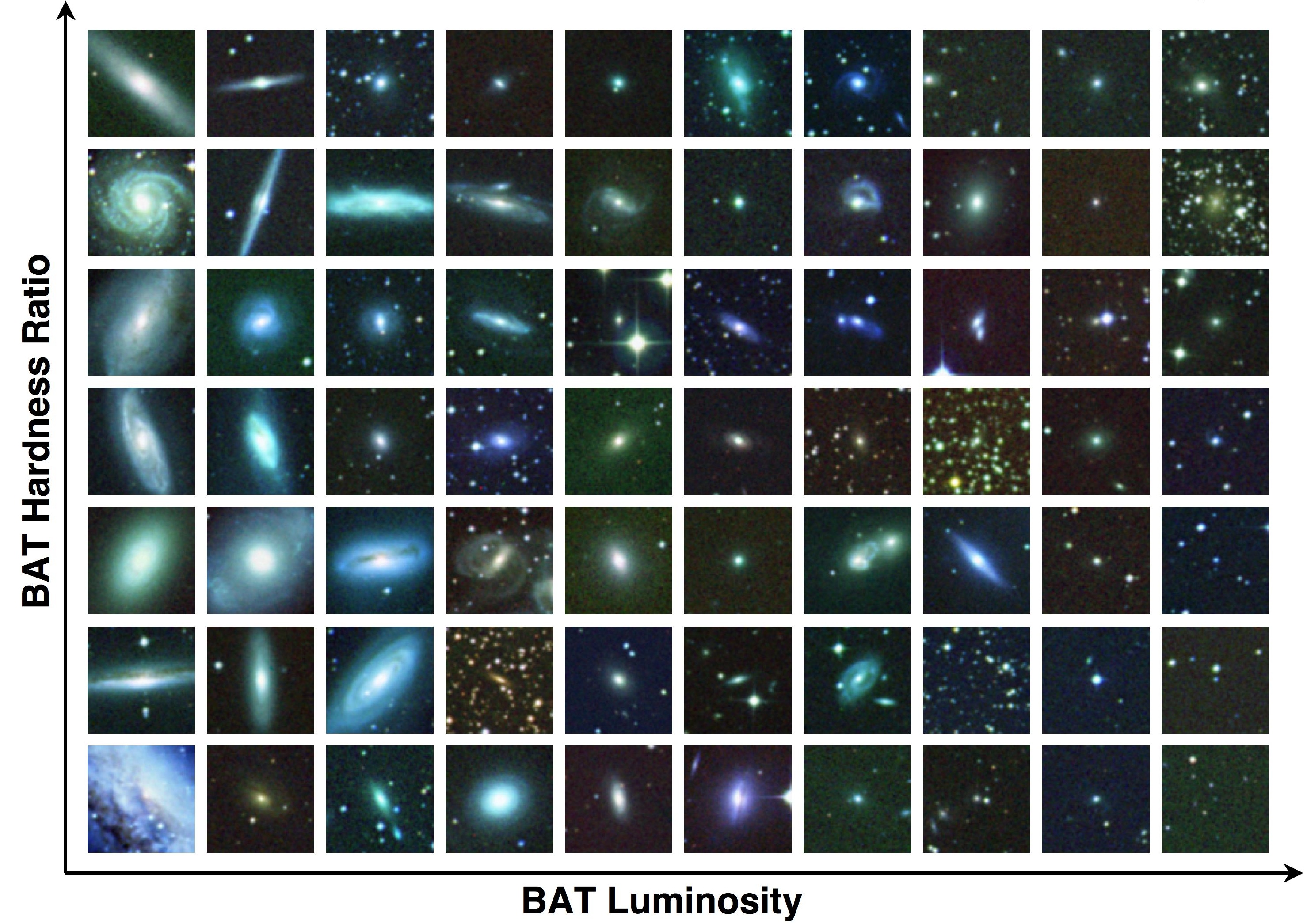}}
\caption{Typical host galaxies of BAT detected Seyfert galaxies.  The
  individual images are taken from the Palomar Digital Sky Survey, and are
  2\arcmin\ on a side.  The BAT hardness-luminosity plane is divided
  into 70 bins, and a BAT source from that bin randomly selected to
  display.
  \label{pretty_pictures}}
\end{center}
\end{figure*}

Figure~\ref{pretty_pictures} shows some typical BAT source host galaxy
images from the Palomar digital sky survey.  The field of view is
2~arcmin across for each subimage.  The figure was produced by
dividing the hardness-luminosity plane into 70 bins and randomly
choosing a BAT source from that category to display.  Of note are the
high fraction of spiral galaxy hosts (as opposed to ellipticals), and
the high number of interacting galaxies.

The 234 sources that have an identification with a well established
Seyfert galaxy more than double the number in any previous hard X-ray
survey. The distribution of column densities, spectral indices, and
luminosities for the survey sources will be presented in a separate
paper. As elsewhere in this paper, we emphasize that this catalog is
based on mean flux levels over the entire 22 month period.  The
detections of sources with significant temporal variability over the
survey period and the implications of such variability will be
discussed elsewhere (Skinner et al., in preparation).

\begin{figure*}
\begin{center}
\resizebox{0.48\textwidth}{!}{\rotatebox{180}{
    \includegraphics{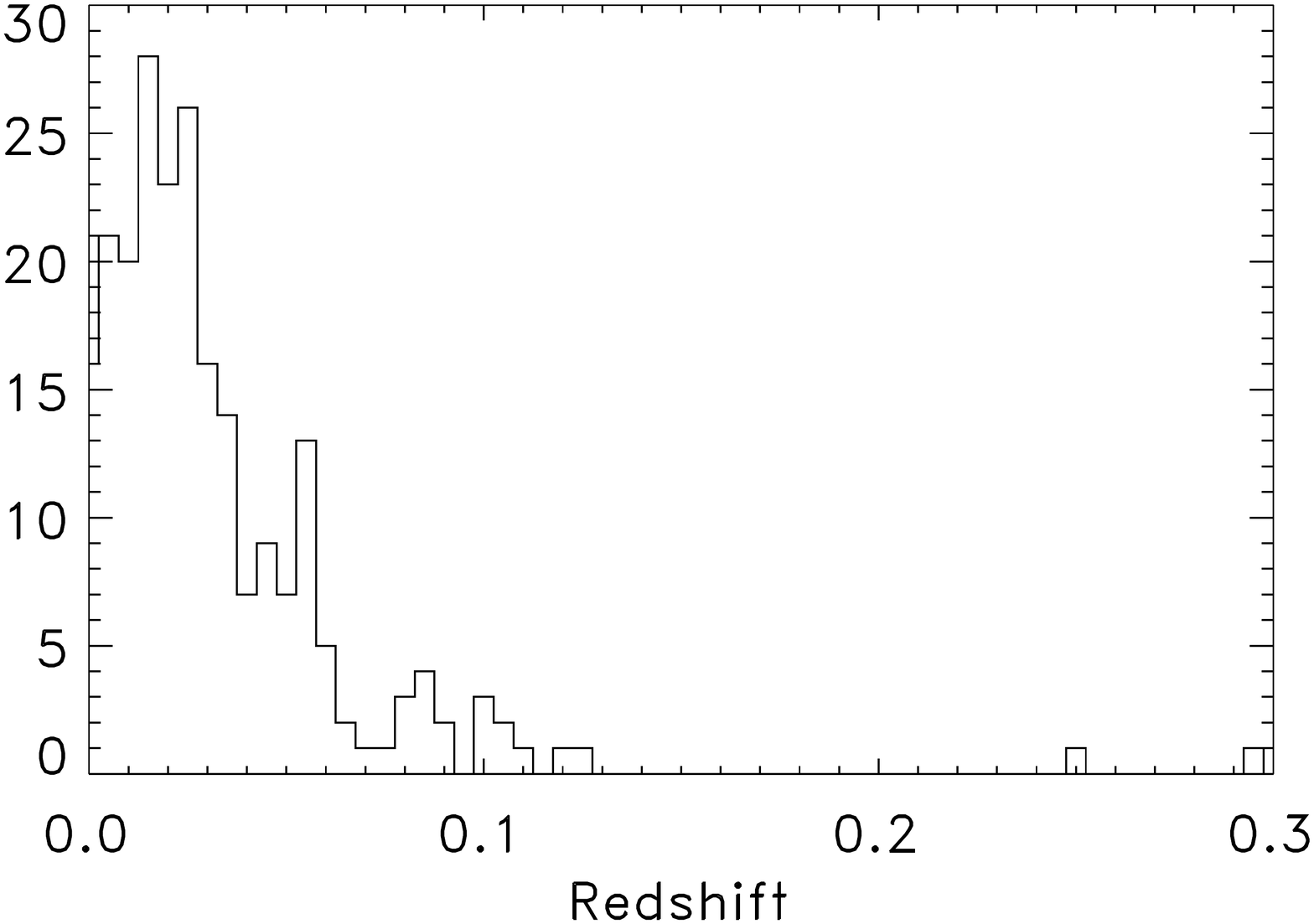}}}
\resizebox{0.48\textwidth}{!}{\rotatebox{180}{
    \includegraphics{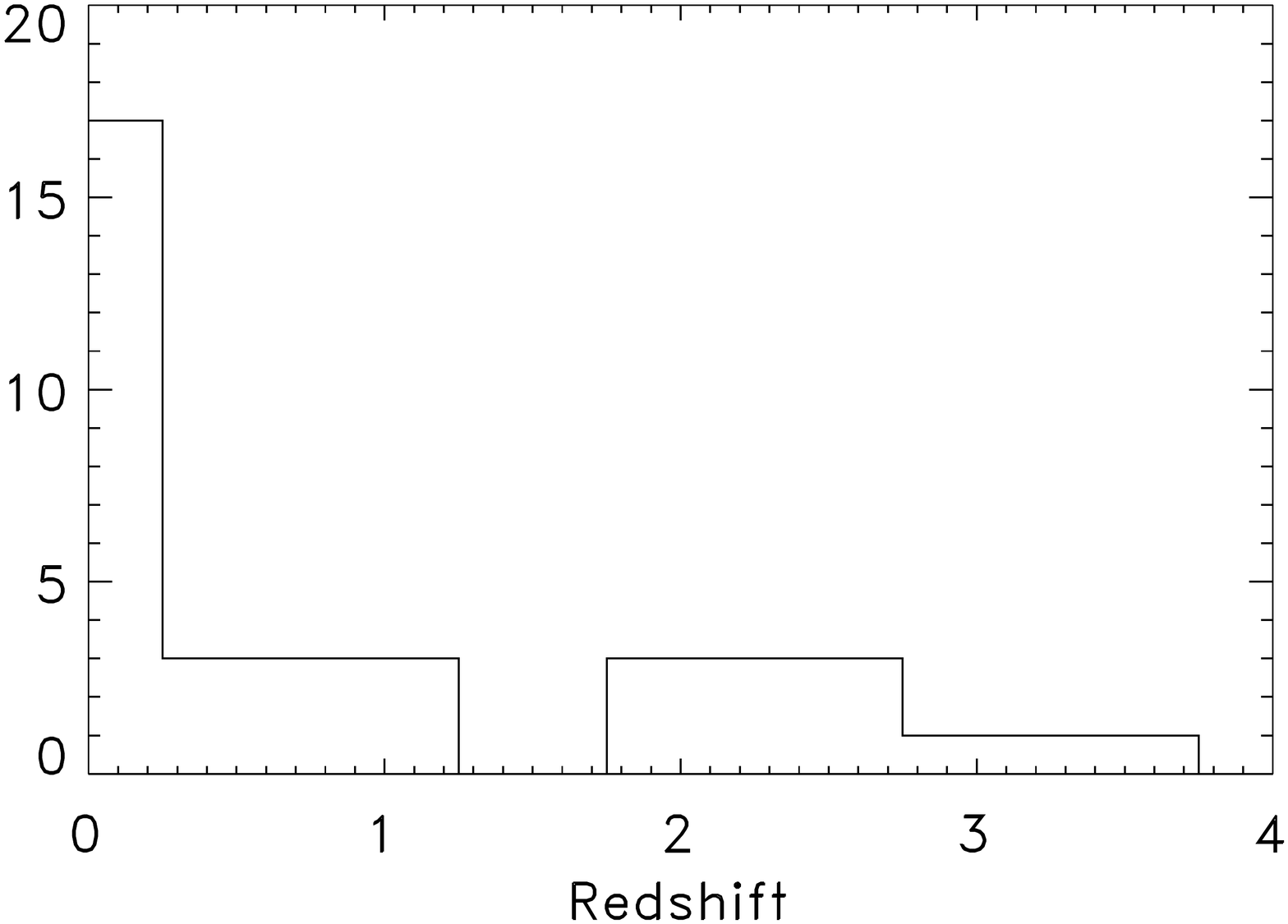}}}
\caption{Histograms showing the redshift distribution of the AGN in the
  22-month survey.  The left panel shows the Seyfert distribution, and
  the right panel shows the beamed AGN distribution.  The beamed AGN
  panel has redshift bins that are 0.5 units wide.
\label{redshift_histo}}
\end{center}
\end{figure*}

Figure~\ref{redshift_histo} shows a histogram of the redshifts of all
the AGN found in the 22-month catalog.
The distribution of the Seyfert galaxy redshifts from the 22-month
survey (left panel of Figure~\ref{redshift_histo}) is highly biased
towards low redshifts ($z\sim 0.03$) with a tail extending out to
$z\sim 0.1$ and a few more distant objects out to $z \sim 0.3$. The
right panel of Figure~\ref{redshift_histo} shows the redshift
distribution of the beamed AGN.  This distribution is quite different
from the Seyfert galaxies in the left panel, with redshifts that
extend to $z\sim 4$ and with no objects at $z<0.033$. Since we have no
selection biases with respect to these beamed AGN (as opposed to
optical searches for blazars) these different redshift distributions
are a fundamental property of these classes and are directly related
to their luminosity functions and evolution \citep{ajello09b}.

\begin{figure}[t]
\begin{center}
\resizebox{\columnwidth}{!}{\includegraphics{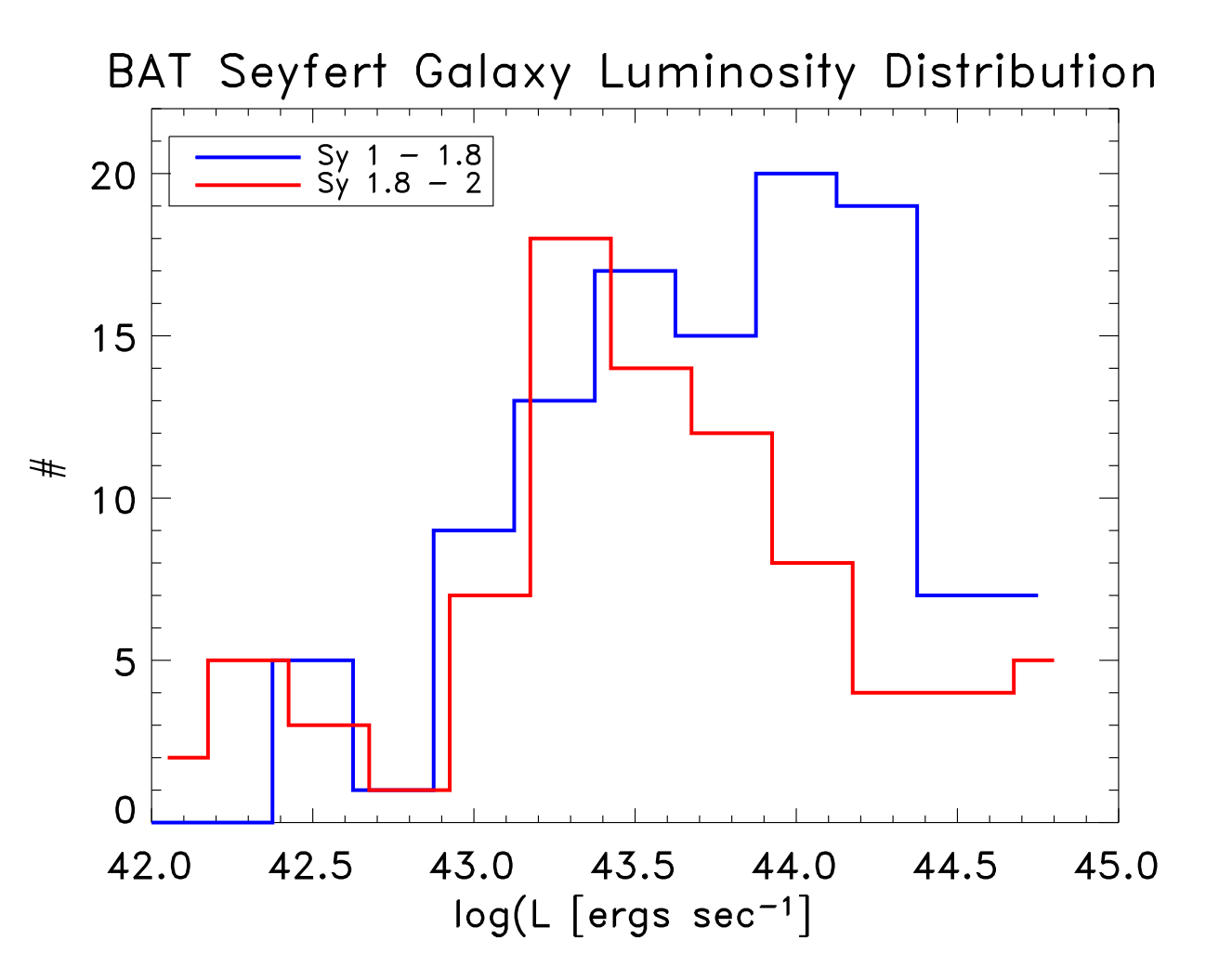}}
\caption{The distribution of Seyfert galaxy luminosities in the BAT
  22-month survey.  \label{lum_dist}}
\end{center}
\end{figure}

Figure~\ref{lum_dist} is a histogram of the luminosities of the
Seyfert galaxies detected in the BAT 22-month survey.
The luminosity distribution of the Seyfert galaxies continues to show
a difference between the type Is and type IIs as noted in
\cite{winter08}. This indicates that the true luminosity distribution
is indeed different for these two classes, which is inconsistent with
the unified model of AGN. A K-S test of these two luminosity
distributions shows that they are drawn from the same parent
distribution with a probability of only 0.30.

As in earlier hard X-ray surveys, the second most common category of
extragalactic sources are beamed AGN, which include types such as
blazars, BL~Lacs, FSRQs, etc. There are 32 objects in this category.
The highest redshift is for QSO~J0539$-$2839 at $z=3.104$.

We have detected 10 clusters of galaxies at $>4.8\sigma$; Perseus,
Coma, Ophiuchus, Cygnus~A, Abell~2319, Abell~754, Abell~3266,
Abell~2142, Abell~3571, and Triangulum~Australis. The BAT spectra of
Cygnus~A and Abell~2142 in the 14--100 keV band are dominated by the AGN
component in or around the clusters. The other 8 clusters are all hot
($kT\sim$10 keV); their BAT spectra are consistent with an extension
of the thermal emission modeled with \asca/\xmm/\chandra\ archival
data in the 2--10 keV band and do not require any additional component
for a good fit. In other words, there is no evidence of a non-thermal
diffuse component in these clusters. We estimated the upper limit of
the non-thermal emission by adding a power-law model to the spectral
fit for the 10 detected clusters. The upper limit is $\sim$
6$\times$10$^{-12}$ ergs cm$^{-2}$ s$^{-1}$ on average and the lower
limit for the magnetic field $B$ ranges from $\sim 0.2$--1~$\mu$G,
assuming inverse Compton scattering of Cosmic Microwave Background
photons by relativistic electrons in the cluster. More details are
described in the papers by \cite{ajello09} and Okajima \etal (ApJ,
submitted).

Some 20 sources are clearly identified, largely through follow-up
\swift-XRT observations, with galaxies from which no sign of nuclear
activity has yet been reported in other wavebands. Their mean
luminosity is only slightly lower than that of those classed as
Seyferts on the basis of optical spectra (10$^{43.53}$ compared with
10$^{43.75}$ erg s$^{-1}$).  These are probably low-$z$ counterparts
to the X-ray bright, optically normal (XBONG) sources discovered in
the \chandra\ and \xmm\ deep fields \citep{barger05,comastri02}.

\subsection{Galactic Sources}

\subsubsection{X-ray Binaries}

As can be seen from Table~\ref{table_sources}, approximately
two-thirds of the Galactic sources are X-ray binaries.  Of those whose
nature is known, about 40\% are high mass X-ray binaries, which
reflects the BAT's sensitivity to hard-spectrum X-ray sources. 60\%
are low mass X-ray binaries, which typically have softer spectra but
can have a higher total flux.  The low mass X-ray binary population is
concentrated near the Galactic plane and bulge, whereas the high mass
X-ray binary population is more distributed, including significant
contributions from the Magellanic clouds.

The sensitivity of the survey is such that high luminosity sources
($>L_x\sim10^{36}$~erg~s$^{-1}$) are detectable anywhere in the Galaxy
and the catalog is complete for sources that emit continuously at this
level. However, since many X-ray binaries are transient it is likely
that there are a significant number of additional X-ray binaries that
are not seen in the present analysis (which is based on fluxes
averaged over 22 months), but that can be detected in specific shorter
intervals. This will be the subject of further work.  The detection of
outbursts from transients that do not repeat, or that repeat only on
timescales of several years, should scale approximately linearly with
the length of the survey. 

\subsubsection{CVs and Other Accreting White Dwarf Systems}

Accreting white dwarf binaries constitute the second most common
category of Galactic sources.  Of these, 31 have been identified as
cataclysmic variables (CVs) or CV candidates.  Of the 31 CVs detected
with BAT, 14 are in the \cite{barlow06} (\integral) list, while 17 are
new detections in hard X-ray surveys.  Because the \integral\ catalog
goes deeper near the Galactic plane but the BAT catalog is more
sensitive at higher latitude, the lists of CVs in the two catalogs are
complementary.  With the expanded list, we confirm that the hard X-ray
selected CVs are dominated by magnetic CVs of the intermediate polar
(IP) subtype, also known as DQ~Her type stars (see also \cite{brun}).

In addition, 4 hard X-ray bright symbiotic stars have been detected by
BAT, as summarized by \cite{kennea07}, and there is now a
candidate to be the fifth member of the class.  Finally, BAT has also
detected one Be~star, gam~Cas, for which an accreting white dwarf
companion is one of three possibilities proposed as the origin of the
X-ray emission \citep{kubo98}.

\subsubsection{SNRs and Non-Accretion Powered Pulsars} 

We detect hard X-ray emission from 8 pulsars and/or their associated
Pulsar Wind Nebula (PWN) or supernova remnants (SNRs).  Our upper
limit on PSR J1846$-$0258 is consistent with the flux level at which
it was detected in a long \integral-IBIS observation and reported in
\cite{3rdBird}. In the case of HESS~J1813$-$178 it appears that we are
detecting emission that is from the point source seen at lower
energies (\cite{funk07}) rather than directly related to the slightly
extended VHE gamma-ray emission. The only SNR-related source that is
not associated with a PWN is Cas A. 

\subsubsection{The Galactic Center} 

Because of the limited resolution of the BAT instrument the emission
reported as from Sgr~A* should be regarded as the net emission from a
region of $\sim 6$\arcmin\ radius centered on the Galactic center. It
is possible that a number of sources contribute.

%----------------------------------------------------------------------------
\section {Conclusions}

The 22 month BAT catalog reinforces and enhances the results from the
3 month \citep{markwardt05} and 9 month \citep{tueller9} catalogs and
shows that the BAT survey continues to increase in sensitivity roughly
as the square root of time and is far from being confusion or
systematics limited. Future papers will discuss the X-ray spectral and
timing properties of these sources as well as the nature of the
optical identifications, in a manner similar to that of
\cite{winter08,winter09}. We have also obtained extensive optical
imaging and spectroscopy of the AGN population (Koss, in preparation;
\cite{winter09}), more detailed X-ray observations as well as Spitzer
IR spectroscopy \citep{melendez09} and radio data which will be
presented in future publications. These results will allow the first
large scale categorization of the AGN phenomenon from a large uniform
and unbiased sample as well as a detailed comparison with results
obtained with other selection techniques.

%--------------------------------------------------------------------------

%% Stuff needed to get big table to go correctly (see emulateapj.cls)
\clearpage
\LongTables
\begin{landscape}

\tabletypesize{\scriptsize}
\setlength{\tabcolsep}{0.035in}
\renewcommand{\arraystretch}{1.05}
\begin{deluxetable}{rrrrrllrrrrrrrrrl}
\tablecaption{Catalog of Sources in the 22-month \swiftbat\ Survey \label{table_sources}}
%\rotate
\tablewidth{0pt}
\tablehead{
\colhead{\#}& \colhead{BAT Name\tablenotemark{a}}& \colhead{RA\tablenotemark{b}}& \colhead{Dec}& \colhead{S/N}& \colhead{Counterpart Name}& \colhead{Other Name} & \colhead{Ctpt RA\tablenotemark{c}} & \colhead{Ctpt Dec} & \colhead{Flux\tablenotemark{d}} &\colhead{error} &\colhead{C\tablenotemark{e}} &\colhead{Hrat\tablenotemark{f}} &\colhead{Herr} &\colhead{Redshift\tablenotemark{g}} & \colhead{Lum\tablenotemark{h}} & \colhead{Type}
}
\startdata
  1&SWIFT J0006.2$+$2012&  1.555& 20.192&   7.05&MRK 0335                 &                  &  1.5813&  20.2029&   2.47&0.41& &  0.31&   0.07&0.0258&43.57&Sy1.2          \\
  2&SWIFT J0010.5$+$1057&  2.613& 10.958&   7.59&MRK 1501                 &                  &  2.6292&  10.9749&   4.10&0.55& &  0.40&   0.05&0.0893&44.91&Sy1.2          \\
  3&SWIFT J0025.8$+$6818&  6.443& 68.303&   5.76&2MASX J00253292+6821442  &                  &  6.3870&  68.3623&   3.02&0.66& &  0.51&   0.09&0.0120&42.99&Sy2            \\
  4&SWIFT J0026.1$+$0508&  6.550&  5.121&   5.55&SWIFT J002615.1+050417   &                  &  6.5631&   5.0715&   4.13&1.12& &  0.59&   0.12&     ?&     &               \\
  5&SWIFT J0028.9$+$5917&  7.224& 59.284&  24.30&V709 Cas                 &RX J0028.8+5917   &  7.2036&  59.2894&   8.91&0.39& &  0.25&   0.02&      &     &CV/DQ Her      \\
  6&SWIFT J0036.0$+$5951&  9.000& 59.842&   5.76&QSO B0033+595            &1ES 0033+595      &  8.9694&  59.8346&   1.96&0.34& &  0.19&   0.07&0.0860&44.56&BL Lac         \\
  7&SWIFT J0037.2$+$6123&  9.297& 61.380&   6.45&BD +60 73                &IGR J0370+6122    &  9.2902&  61.3601&   2.85&0.47& &  0.35&   0.06&      &     &HMXB           \\
  8&SWIFT J0042.9$-$2332& 10.745&$-$23.503&   7.00&NGC 235A                 &                  & 10.7200&$-$23.5410&   4.12&0.48& &  0.35&   0.05&0.0222&43.67&Sy1            \\
\enddata
\tablenotetext{a}{If a BAT name exists in the 9-month catalog \citep{tueller9}, then that name is used.  If there is no 9-month BAT name, then the BAT name listed here is the name that was used to request XRT followup observations (and used in the HEASARC archive).  When no previous BAT name for this source exists, we list here a BAT name derived from the BAT position in this catalog.}
\tablenotetext{b}{The BAT source positions listed here are all uniformly generated from a blind search of the 22-month data and are J2000 coordinates.}
\tablenotetext{c}{The counterpart position is the most accurate position known of the object in the 'name' column in J2000 coordinates, and is usually taken from NED or SIMBAD.  If no counterpart is known, the blind BAT position is listed.}
\tablenotetext{d}{The flux is extracted from the BAT maps at the position listed for the counterpart, is in units of $10^{-11}$ergs\,s$^{-1}$cm$^{-2}$, and is computed for the 14--195~keV band.}
\tablenotetext{e}{The meaning of the confusion flag is as follows: A = 'Confused' source, B = 'Confusing' source, AB = Both confused and confusing.  To make sense of this confounding scenario, see \S\ref{catalog} and \S\ref{confused}.}
\tablenotetext{f}{The hardness ratio is the ratio of the 35--150~keV count rate to the 14--150~keV count rate.}
\tablenotetext{g}{The redshifts are taken from the online databases NED and SIMBAD or in a few cases from our own analysis of the optical data.  A blank indicates that the object is Galacatic, and a ? indicates that the object has an unknown redshift.}
\tablenotetext{h}{The luminosity is computed from the flux and redshift in this table, with units of log[ergs\,s$^{-1}$] in the 14--195~keV band.}
\tablerefs{\cite{bikmaev08,bikmaev06,bodaghee06,bodaghee07,burenin08,butler09,chakrabarty02,coe94,homer02,israel01,juett05,markwardt05,masetti08,masetti09,rodriguez09,tomsick08,tueller9,walter06,wilson03}}
\end{deluxetable}
\renewcommand{\arraystretch}{1.0}

\clearpage
\end{landscape}

\end{document}